\documentclass[11pt]{article}
\usepackage[margin=1in]{geometry}
\usepackage{authblk}
\usepackage[table,dvipsnames]{xcolor}
\definecolor{NCSUred}{RGB}{204,0,0}
\definecolor{NCSUyellow}{RGB}{250,200,0}
\definecolor{NCSUorange}{RGB}{209,73,5}
\definecolor{NCSUdarkred}{RGB}{153,0,0}
\definecolor{NCSUaqua}{RGB}{0,132,115}
\definecolor{NCSUgreen}{RGB}{111,125,28}
\definecolor{NCSUblue}{RGB}{65,86,161}
\definecolor{NCSUlightblue}{RGB}{66,126,147} 
\usepackage{graphicx, subcaption, url}
\usepackage[colorlinks=true, linkcolor=NCSUblue, urlcolor=NCSUaqua, citecolor=NCSUblue]{hyperref}
\usepackage{amsmath, amsfonts, amssymb, amsthm}
\allowdisplaybreaks
\theoremstyle{plain}
\newtheorem{remark}{Remark}[section]
\newtheorem{assumption}{Assumption}[section]
\newtheorem{definition}{Definition}[section]

\newtheorem{proposition}{Proposition}[section]

\newtheorem{theorem}{Theorem}[section]

\newtheorem{fact}{Fact}[section]

\RequirePackage[T1]{fontenc}
\usepackage{stix2}
\usepackage[cal=rsfso, bb=ams]{mathalpha}
\usepackage[per-mode=repeated-symbol]{siunitx}

\newcommand{\mbb}[1]{\mathbb{#1}}
\newcommand{\mc}[1]{\mathcal{#1}}
\newcommand{\mr}[1]{\mathrm{#1}}
\newcommand{\ms}[1]{\mathsf{#1}}

\newcommand{\dif}[1]{\mathrm{d} #1}
\newcommand{\Dif}[2]{\mathrm{D}^{#2} #1}

\newcommand{\mE}{\mathbb{E}}

\newcommand{\mN}{\mathbb{N}}
\newcommand{\mP}{\mathbb{P}}
\newcommand{\mR}{\mathbb{R}}
\newcommand{\mst}[1]{\mathrm{\, s.t. \,}}

\newcommand{\mgk}{\kappa}
\newcommand{\mgvk}{\varkappa}
\newcommand{\mRd}[1]{\mathbb{R}^{#1}}

\newcommand{\rg}[1]{\mathring{#1}}
\newcommand{\mCb}{C_{\mathrm{b}}}
\newcommand{\rCb}{\rg{C}_{\mathrm{b}}}

\newcommand{\bra}[1]{\left(#1\right)}
\newcommand{\Bra}[1]{\left[#1\right]}
\newcommand{\BRA}[1]{\left\{#1\right\}}
\newcommand{\ip}[2]{\left\langle #1 , #2 \right\rangle}
\newcommand{\norm}[1]{\left\Vert #1 \right\Vert}

\title{\bf Koopman–Nemytskii Operator of Nonlinear Controlled
Systems and Its Learning for Controller Synthesis
\thanks{This paper was submitted to \textit{IEEE Transactions on Automatic Control} in its third edition on June 17, 2026. The previous two versions, submitted on March 24, 2025 and October 3, 2025, had the title ``\textit{Koopman-Nemytskii Operator: A Linear Representation of Nonlinear Controlled Systems}''. This work is supported by NSF (CBET Award \#2414369). The codes are available at the author's GitHub Repository: \url{https://github.com/WentaoTang-Pack/Koopman-Nemytskii}.}}
\author[1]{Wentao Tang \thanks{Corresponding author: \href{mailto:wtang23@ncsu.edu}{\tt wtang23@ncsu.edu} }}
\affil[1]{Department of Chemical and Biomolecular Engineering, North Carolina State University}
\date{June 17, 2026}

\begin{document}
\pagenumbering{roman}
\maketitle
\pagenumbering{arabic}

\begin{abstract}
While the Koopman operator represents a nonlinear system as a linear operator in a function space, its definition does not involve inputs. 
For controller synthesis, an operator model is needed to describe the effect of feedback laws on closed-loop systems, so that the desired state-feedback law can be computationally searched based on such a predictive model. 
To this end, this paper proposes a Koopman--Nemytskii operator, defined as a linear operator that maps canonical features of state--policy pairs in a reproducing kernel Hilbert space (RKHS) to that of succeeding states. 
Under regularity conditions on the dynamics and kernel selection, this operator is definable on suitable Sobolev-type RKHSs, and its data-based estimation guarantees bounded errors in single-step prediction, multi-step prediction, and accumulated cost under control. 
The controller synthesis problem is thus formulated as a convex kernel-based optimization one and efficiently solved in a sample-based manner. 
\end{abstract}

\section{Introduction}\label{sec:Introduction}
Linearization is a fundamental idea in nonlinear control, such as feedback linearization \cite{baillieul1999nonlinear}, input--output linearization \cite{isidori1985nonlinear}, Carleman linearization \cite{kowalski1991nonlinear}, and the ``Koopmanist'' framework that has received extensive studies recently \cite{mauroy2020koopman, mezic2021koopman, brunton2022modern}.
\emph{Koopman operator} \cite{koopman1931hamiltonian} is a representation of nonlinear dynamics in a generically \emph{infinite-dimensional function space} as a linear mapping. 
In such a Koopman framework, many classical nonlinear control problems such as observer design \cite{yi2023equivalence, ye2025edmd, tang2025data} and optimal control \cite{moyalan2023data, hoischen2025operator} can be potentially reformulated in data-driven manners as convex problems. 

Specifically, for a discrete-time system:
\begin{equation}\label{eq:autonomous.system}
	x_{t+1} = f(x_t), \, x_t\in \mbb X \subset \mRd{d_x}, \, t=0,1,2,\dots, 
\end{equation}
the Koopman operator $K_f$ is the linear operator on a linear space of state-dependent functions $\mathcal{G}$, given by 
\begin{equation}\label{eq:Koopman.operator}
	K_fg = g\circ f, \enskip g\in \mc{G},
\end{equation}
i.e., $(K_fg)(x)=g(f(x))$ ($\forall x\in \mbb X$). 
However, its extension to systems with inputs (i.e., \emph{controlled} or \emph{actuated} systems):
\begin{equation}\label{eq:controlled.system}
	x_{t+1} = f(x_t, a_t), \, x_t\in \mbb X\subset \mRd{d_x}, \, a_t\in \mbb A\subset \mRd{d_a}, 
\end{equation}
where $a_t$ is the inputs (actions), is nontrivial. 
It was often assumed (without enough rigor) that the open-loop dynamics of \eqref{eq:controlled.system} can be approximated as a linear or bilinear one in a lifted but still finite-dimensional space, e.g., in \cite{korda2018linear, huang2022convex, strasser2024koopman}. 
In Williams et al. \cite{williams2016extending}, it was first proposed that the Koopman operator for controlled systems can be (approximately) considered as multiple Koopman operators parameterized by \emph{input values} $a_1, \dots, a_{d_a}$, namely $K:= K_0 + \sum_{j=1}^{d_a} a_j K_j$. Such a combination is exact if the system is continuous-time and the Koopman operators are replaced by the infinitesimal generators of corresponding semigroups \cite{goswami2021bilinearization}. 
More generally, the Koopman operator can be nonlinearly parameterized by input values \cite{peitz2020data, bonnet2024set} or even input sequences \cite{haseli2025koopman}. 

\par The concept of \emph{reproducing kernel Hilbert spaces} (RKHSs) provides more rigorous operator models for \eqref{eq:controlled.system}. 
Bevanda et al. \cite{bevanda2024nonparametric} proposed a ``control Koopman operator'' from $L^2(\mbb X\times \mbb A)$ to $L^2(\mbb X)$, approximated it on an RKHS, and discussed its learning as a Hilbert--Schmidt operator (with technical analysis similar to autonomous systems  \cite{kostic2022learning}). 
The idea of kernel-based Koopman model was used in Str{\"{a}}sser et al. \cite{strasser2025kernel} and Bold et al. \cite{bold2025kernel} for analyzing Koopman control of continuous-time input-affine systems. 
In Hou et al. \cite{hou2026behavioral}, kernel-based Koopman models for discrete-time Volterra and Hammerstein systems were used for their control in behavioral frameworks. 
For discrete-time systems, the work of Lazar \cite{lazar2025product} 
defined the Koopman operator on the tensor product of a state-RKHS and an input-RKHS. 
In the author's recent work \cite{morris-ye-tang2026}, it was proved that under regularity assumptions on the dynamics \eqref{eq:controlled.system}, by lifting state and input onto Sobolev-type RKHSs, an operator model that involves state feature and input feature \emph{bilinearly} indeed holds exactly in discrete time. 

\par Essentially, the existing Koopman operator models for controlled systems aim to account for the effect of \emph{open-loop input values}, i.e., to establish an operator $O$ that maps the state feature ($\phi$) and input feature ($\phi'$) to the succeeding state feature:
\begin{equation}\label{eq:conceptual.input-state.viewpoint}
    O: (\phi_{x_t}, \phi_{a_t}') \mapsto \phi_{x_{t+1}}. 
\end{equation}
Hence, the learning of such a relation, embodying the idea that ``similar input values cause similar transitions'', is suitable for \emph{open-loop} prediction and optimization of \emph{open-loop} control schedules, e.g., in model predictive control (MPC) \cite{korda2018linear, bevanda2024koopman, bold2025kernel}. 
However, when concerned with \emph{controller synthesis}, one tends to be interested in optimizing an explicit \emph{feedback law} or \emph{policy}: $a = u(x)$, so that the closed-loop system:
\begin{equation}
	x_{t+1} = f(x_t, u(x_t)) =: f_u(x_t)
\end{equation}
achieves stability and performance specifications. 
Hence in this paper, an operator $T$ is sought to map the state feature ($\phi$) and some policy feature ($\varphi$) to the succeeding state feature:
\begin{equation}\label{eq:conceptual.policy-state.viewpoint}
    T: (\phi_{x_t}, \varphi_{u_t}) \mapsto \phi_{x_{t+1}}. 
\end{equation}

\subsection{Motivation for a policy--state operator model: Equilibrium and stability}\label{subsec:motivation.1}
The first justification for using an operator \eqref{eq:conceptual.policy-state.viewpoint} that embodies the policy--state viewpoint, instead of the input--state viewpoint as in \eqref{eq:conceptual.input-state.viewpoint}, is the conceptual suitability for closed-loop control. 
Let us consider a continuous-time system\footnote{We may discretize the time by a small constant, without changing the qualitative observations in this subsection.}:
\begin{equation}
	\mr{d}x_t/\mr{d}t = x_t - 2x_t^2\mr{sgn}(x_t) + a_t
\end{equation}
where $a_t\in \mbb A=[-1,1]$. The system has an invariant set $\mbb X=[-1, 1]$ under the given $\mbb A$. One can easily verify that when $a=\pm 1$, the state is attracted to $\pm1$, respectively, and that when $a=0$, the state has three equilibrium points: $0$, $1/2$, and $-1/2$, among which the origin is unstable and the latter two are asymptotically stable. 
Although the flow (vector field) under $a=0$ is indeed the average of the flows under $a_t\equiv\pm1$, the behavior of the system changes qualitatively. 

\par
\begin{figure}[!t]
	\begin{center}
		\includegraphics[width=0.75\columnwidth]{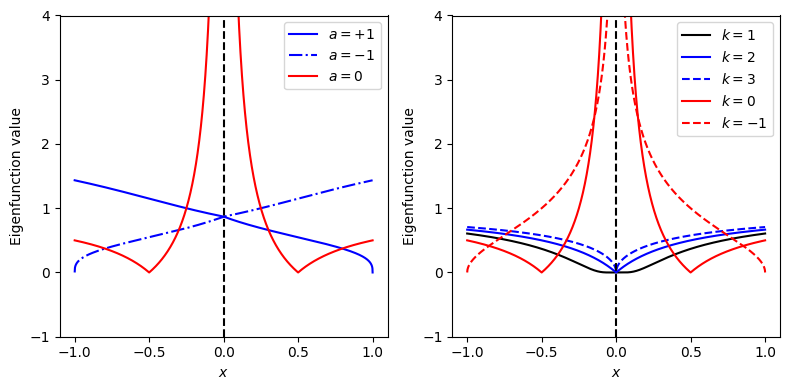}
		\caption{Koopman eigenfunctions under different open-loop input values (left) versus under different feedback gains (right).}
		\label{fig:landau_stuart}
	\end{center}
	\vspace{-1.5em}
\end{figure}

 If one has the input--state operator model \eqref{eq:conceptual.input-state.viewpoint} exactly, then the Koopman eigenfunctions\footnote{That is, a function whose value decays at the rate of $e^{\lambda t}$ in continuous time. Here we consider $\lambda = -1$, and formally allow an eigenfunction to take value on the extended real line $\mR\cup\{\infty\}$ to make it well-defined.} appear as in the left subplot of Fig. \ref{fig:landau_stuart}. 
Clearly, the eigenfunction at $a=0$ by no means resemble those at $a=\pm1$ or any combination of them. Hence, the information of equilibrium points and their stability, in principle, may not be readily ``interpolated'' from the dynamics under different input values. 
When one aims to obtain a controller that stabilizes the origin, it may not be a most natural idea to achieve closed-loop stability by a scheduled input sequence.\footnote{It should be clarified that although the input schedule can be optimized by MPC, and the stability can be achieved under suitable conditions on the operator modeling error \cite{bold2025kernel}, the problem of synthesizing stage cost and terminal cost functions in MPC (which implicitly contains an underlying controller synthesis problem) were essentially unanswered in the Koopman-based MPC approach.} 

\par In contrast, given a suitable family of feedback laws, it is possible to fix the equilibrium point. Consider linear feedback laws: $a_t=u(x_t) = -\mr{sat}(kx_t)$, which stabilize the origin when $k\geq 1$. The resulting eigenfunctions are illustrated in the right subplot of Fig. \ref{fig:landau_stuart}; clearly, the eigenfunctions appear similar at close $k$ values, e.g., when $k=2$ and $k=3$, without losing the information of equilibrium. 
Therefore, for controller synthesis, it is naturally desirable to learn an operator to directly tell that ``similar \emph{feedback laws} shall result in similar \emph{closed-loop dynamics}''. This is exactly what \eqref{eq:conceptual.policy-state.viewpoint} aims for. 

\subsection{Motivation for a policy--state operator model: Controller synthesis}\label{subsec:motivation.2}
\par Consider the problem of finding an optimal controller with minimized cost:
$ J(k)=\mE_{x_0\sim \mathrm{Unif}[-1, 1]} \int_0^\infty (4x_t^2 + a_t^2) \dif{t}$.
By a black-box search over one dimension ($k$), the optimal linear feedback law is found as $k=k^\ast\approx2.34$; clearly, such a search relies on an underlying sufficiently smooth relation between $k$ and $J(k)$. In fact, the controller synthesis always boils down to \emph{policy optimization}, namely the search over the parameters in the feedback law, and all feedback laws in scope are usually restricted in an admissible class. 
As is well-known, for nonlinear systems, especially when the governing equations are unknown, the computation for an optimal controller is generally difficult \cite{lewis2012optimal}. Hence, an operator model that involves feedback policies (instead of involving the open-loop input values) would bring simplification, because with this model, the controller synthesis can be treated as a \emph{direct search over policy features}. It will be shown that its solution can be well approximated in a data-driven (sample-based) manner. 

\par In contrast, suppose that the open-loop input--state operator model \eqref{eq:conceptual.input-state.viewpoint} is used. When the controller is restricted to a parameterized family of laws, e.g., $u(x_t)= \mr{sat}(-kx_t)$, the synthesis problem would query the state prediction by this operator in composition with the controller parameter: $(\phi_{x_t}, \phi_{\mr{sat}(-kx_t)}' )\mapsto \phi_{x_{t+1}}$. 
Still, only the predictions on the state--input pairs that conform to the parameterized laws are unused. 
Hence there is no saving on the information needed for learning the policy--state operator \eqref{eq:conceptual.policy-state.viewpoint}, and in fact, the choice of an input-value kernel $\phi'$ to meaningfully define the composition and relate the policy to performance brings back the notion of \eqref{eq:conceptual.policy-state.viewpoint}. 
On the other hand, if an open-loop model \eqref{eq:conceptual.input-state.viewpoint} is used for an unrestricted controller synthesis, then the resulting solution is difficult due to the requirement of composition with undetermined nonparametric functions.

\subsection{Contributions of this work}
In this paper, we propose a novel concept of \emph{Koopman--Nemytskii operator}, which is a generic linear representation of nonlinear systems in discrete time without input affinity. It characterizes the closed-loop flow under feedback laws from a \emph{policy space} and is useful for controller synthesis as a search over the policy features. 
\begin{enumerate} 
	\item 
    To the end of capturing closed-loop flows with a fixed equilibrium point at the origin, the mathematical construction of Koopman--Nemytskii operator relies on a \emph{linear--Sobolev RKHS} $\mc{N}_\kappa(\mbb{X})$ associated with a \emph{linear--radial kernel} $\kappa$. 
    This ``equilibrium-preserving'' technique was introduced in the author's work on Koopman spectrum--stability relation \cite{tang-ye2025koopman}, extended to Lyapunov and dissipativity analysis \cite{tang-ye2026koopman, tang-ye2026dissipativity}, and used for open-loop bilinear Koopman modeling \cite{morris-ye-tang2026}. 
    Such an RKHS comprises only of state-dependent functions that are ``locally at least linear'' and ``globally Sobolev'', thus suitable for the analysis of dynamics with an equilibrium point. 
    This work first shows that with smooth enough dynamics and under smooth enough policies, the Koopman operator $K_{f_u}$ is a well-defined and bounded linear operator between RKHSs. 
    \item 
    Under regularity assumptions on the policy parameterization, the dependence of the Koopman operator on the policy is continuous (despite nonlinear). Hence, a universal kernel $\mgvk$ is defined on the policy space $\mbb{U}$ (with feature map $\varphi$). The superposition mapping $u\mapsto K_{f_u}$, namely the relation of policy to closed-loop dynamics, then gives rise to a linear operator $T$:  
	\begin{equation}\label{eq:KN.operator.intro}
		T \in \mc{L}(\mc{N}_\kappa(\mbb X) \otimes \mc{N}_{\mgvk}(\mbb U), \mc{N}_\kappa(\mbb X)), \enspace \phi_x \otimes \varphi_u \mapsto \phi_{f_u(x)}, 
	\end{equation}
	which we call as the \emph{Koopman--Nemytskii operator}. Such an operator can be interpreted as a linear description of the evolution of \emph{canonical features} (on the RKHSs) with the nonlinear dynamics, thus enabling prediction of states and state-dependent functions in closed loop. 
    \item 
    Provided a finite dataset of state--policy--successor snapshots $(x_i, u_i, y_i)$ ($y_i=f(x_i, u_i(x_i))$, $i=1,\dots,m$), \emph{a finite-rank approximation of the operator $\hat{T}$ can be learned} via kernel extended dynamic mode decomposition (kernel EDMD). 
    This paper points out that the learned Koopman--Nemytskii operator has bounded generalized state prediction error, bounded multi-step state prediction error, and bounded accumulated cost errors for policy evaluation, which are linearly/quadratically proportional to the norm of initial state. 
    The error is analyzed via the theory of kernel interpolation \cite{wendland2004scattered} using the concept of fill distance of sample points, which thus scales down with increasing sample size. 
    \item 
    Finally, the Koopman--Nemytskii operator is used for controller synthesis. It is shown that the optimization of policy, considered as the search for the optimal policy parameter feature, can be formulated as a \emph{kernel-based optimization} problem and approximated in finite dimensions for efficient data-driven (sample-based) solution. Numerical implementation on a chemical reactor is used to demonstrate the proposed approach in comparison to true model-based MPC, LQR, open-loop Koopman MPC, and open-loop Koopman synthesis strategies. 
\end{enumerate}

\subsection{Related works}\label{subsec:Related}
\paragraph{System identification} In classical literature, identifying linear transfer-function and state-space models from data has been discussed, in open-loop identification (where the system is perturbed by experimental input signals) and closed-loop identification (where the feedback controller is kept on) \cite{hjalmarsson1996model}. 
Nonlinear structures such as Volterra series, kernel, and neural methods have been considered as flexible alternatives \cite{schoukens2019nonlinear}. 
Many recent works, pertaining to physics-informed machine learning, focused on discovering underlying differential equations \cite{brunton2016discovering}. 
In principle, the present work can be considered as a closed-loop nonlinear system identification method that is generically applicable (under mild regularity assumptions). 

\paragraph{Stability considerations in the learning of nonlinear dynamics} 
While there exist a wide range of machine learning algorithms for learning model equations, constraining it to accommodate prior information is desirable. 
In a control context, stability is a primary concern. To the end of enforcing stability, reparameterization of Koopman operator that guarantees Hurwitz or Schur properties has been used \cite{bevanda2022diffeomorphically, fan2022learning}.
Another idea to preserve stability is to collect orbits over long horizons and set the loss metric or kernel to account for long-term errors \cite{lusch2018deep}. 
More general problem settings include the detection of bifurcation from data-driven models \cite{tang2024data}, understanding the spectrum--stability relation in Koopman operator models \cite{tang-ye2025koopman}, and learning the Lyapunov stability/dissipativity \cite{tang-ye2026koopman, tang-ye2026dissipativity}. 

\paragraph{Hilbert space formulations in infinite-dimensional systems and stochastic control theory}
Dynamical systems governed by partial differential equations and delayed differential equations can be conveniently described in infinite-dimensional Banach and Hilbert spaces. The properties of the operators describing the dynamics have been found useful for the analysis of the controllability, observability, and existence of optimal control in classical studies \cite{bensoussan2007representation}.
This is also the case with dynamics governed by stochastic differential equations, interpreted by Fokker--Planck equations that describe the evolution of density functions \cite{bogachev2022fokker} and hence have a functional state space. 
From a computational point of view, convex optimization tools can often be enabled by operator-theoretic models \cite{henrion2020moment, houska2025convex}; this paper shares the same rationale. 

\paragraph{Data-driven controller synthesis} 
Without explicit models, it is possible to design controllers from data with guaranteed specifications. Such data-driven controllers \cite{coulson2019data, van2023informativity} can be found based on the Willems' fundamental lemma, which states that persistently exciting trajectories can fully recover the behavior of linear systems \cite{willems2005note}. 
For nonlinear systems, a multitude of approaches including polynomial approximation, kernel regression, linear parameter-varying embedding, and nonlinearity cancellation have been proposed \cite{martin2023guarantees, de2023learning}. 
Another promising approach is to learn dissipativity from data, which can lead to guaranteed performance even without the information of a complete model \cite{koch2022determining, tang2021dissipativity}. 
The approach in this paper is data-driven and does not involve an explicit first-principles structure, but a nonparametric empirical model.

\subsection{Organization of the paper and notations}
\par The remainder of this paper is organized as follows. In \S\ref{sec:Preliminaries}, the mathematical preliminaries underlying the present paper is provided, after which the construction and properties of the proposed Koopman--Nemytskii operator are presented in \S\ref{sec:Definition}. 
The data-based estimation of the Koopman--Nemytskii operator and its properties, specifically the error bounds on state prediction and accumulated cost prediction, are discussed in \S\ref{sec:Learning} with a numerical example. 
The formulation and algorithm of Koopman--Nemytskii operator-based controller synthesis, along with the application to a chemical reactor, are shown in \S\ref{sec:Experiments}, and conclusions are given in \S\ref{sec:Conclusion}. 
\paragraph*{Notations} 
Throughout this paper, we use lower-case letters for scalars, vectors, and scalar- or vector-valued functions. Capital letters are used for matrices or operators, and sets. Calligraphic letters represent function spaces or operator spaces (e.g., $\mc{L}$ for the space of bounded linear operators). Inner products are denoted as $\ip{\cdot}{\cdot}$, and norms as $|\cdot|$ or $\|\cdot\|$. We use $\mbb{N} = \{0, 1, 2, \dots\}$, $\mbb{R}_+ = [0, \infty)$, and $\mr{D}$ for derivatives or generalized derivatives.

\section{Preliminaries}\label{sec:Preliminaries}
Here we recollect some mathematical facts regarding Hilbert spaces, RKHS, and Koopman operators for later discussions. 

\subsection{Hilbert space and Sobolev--Hilbert space}
The following definition of Hilbert spaces is commonly known (see, e.g., Lax \cite{lax2014functional}). Let $\mc{H}$ be a linear space on the field of real or complex numbers. $\mc{H}$ may not be finite-dimensional. If an \emph{inner product} $\ip{\cdot}{\cdot}$ (a sesquilinear form on two arguments) is defined on $\mc{H}$, then this space is called an inner product space. If further the norm induced by the inner product ($\|h\| = \ip{h}{h}^{1/2}$, $\forall h\in\mc{H}$) makes $\mc{H}$ a complete metric space, then $\mc{H}$ is said to be a \emph{Hilbert space}. 
If such an inner product is not defined, but $\mc{H}$ is normed and complete, then $\mc{H}$ is a Banach space. 
We say that two Hilbert spaces coincide ($\mc{H}_1 \simeq \mc{H}_2$) if $\mc{H}_1 = \mc{H}_2$ as sets and there exist positive constants $c_2\geq c_1>0$, such that $c_1\|h\|_{\mc{H}_1} \leq \|h\|_{\mc{H}_2} \leq c_2\|h\|_{\mc{H}_1}$ for any element $h$. 

\par A typical example of Hilbert space is the following \emph{Sobolev--Hilbert space}. Let $\mbb{X}\subset \mRd{n}$ be nonempty and $H^0(\mbb{X}) = L^2(\mbb{X})$. First, $H^0(\mbb{X})$ is a Hilbert space with inner product $\ip{h_1}{h_2} = \int_\mbb{X} h_1h_2 \dif{x}$. Then, for $s\in \mbb{N}$, let $H^s(\mbb{X})$ be the space of functions whose weak derivatives up to degree $s$ exist and belong to $L^2(\mbb{X})$.\footnote{
The weak derivative of $h$ with multi-index $\alpha=(\alpha_1, \dots, \alpha_n)$, whose degree is $|\alpha|:=\alpha_1 + \dots + \alpha_n$, refers to the function $\Dif{h}{\alpha}$ satisfying 
$$\int_\mbb{X} \phi \Dif{h}{\alpha} \dif{x} = (-1)^{|\alpha|} \int_\mbb{X} h \frac{\partial^{|\alpha|} \phi}{\partial x_1^{\alpha_1} \dots \partial x_n^{\alpha_n}} \dif{x}, $$
for all $\phi$ that are infinitely smooth and compactly supported in $\mbb{X}$.} 
Then $\mc{H}^s(\mbb{X})$ is a Hilbert space with inner product $\ip{h_1}{h_2} = \sum_{|\alpha|\leq s} {\ip{\Dif{h_1}{\alpha}}{\Dif{h_2}{\alpha}}}_{\mc{H}^0}$. The definition can be extended to all $s\in \mR_+$.\footnote{
For fractional $s = \lfloor s \rfloor + r$, $\lfloor s\rfloor\in \mbb{N}$, $r\in(0, 1)$, the norm is defined as: 
$$\|h\|_{\mc{H}^s}^2 = \sum_{|\alpha|\leq \lfloor s \rfloor} \|\Dif{h}{\alpha}\|_{\mc{H}^0}^2 + \sum_{|\alpha|= \lfloor s\rfloor} \iint_{\mbb{X}^2} \frac{|\Dif{h}{\alpha}(x) - \Dif{h}{\alpha}(y)|^2}{|x-y|^{n+2r}} \dif{x}\dif{y}.$$
In fact, $H^s$ is an \emph{interpolation space} between $H^{\lfloor s\rfloor}(\mbb{X})$ and $H^{\lceil s\rceil}(\mbb{X})$. }
A Sobolev--Hilbert space on $\mbb{X}= \mRd{n}$ can be equivalently defined as the space comprising of functions $h$ whose Fourier transform $\hat{h}$ satisfies $c_1(1+|\omega|^2)^{-s} \leq |\hat{h}(\omega)|\leq c_2(1+|\omega|^2)^{-s}$ for all $\omega\in \mRd{n}$ (with constants $0<c_1\leq c_2$) \cite{leoni2024first}. In other words, the space $H^s$ defined by Fourier transform and the $H^s$ defined by generalized derivatives coincide. 

\par Given two Hilbert spaces $\mc{G}$ and $\mc{H}$, their \emph{tensor product} is 
$$\mc{G}\otimes \mc{H} = \overline{\mr{span}}\BRA{ g\otimes h: g\in \mc{G}, h\in \mc{H}}$$
where the elementary tensors $g\otimes h$, called the \emph{Kronecker product} of $g$ and $h$, are the ordered pairs of elements $(g, h)$ endowed with inner product: $\ip{g_1\otimes h_1}{g_2\otimes h_2} = \ip{g_1}{g_2} \ip{h_1}{h_2}$. 
For clarity, a different notation $g\times h$ is used to refer to the rank-$1$ operator from $\mc{H}$ to $\mc{G}$, defined by $$(g\times h)h' = \ip{h}{h'}g, \enspace \forall h'\in \mc{H}.$$

\subsection{Reproducing kernel Hilbert space (RKHS)}
\par The concept of RKHS is useful for regression or classification tasks. Without prior knowledge for defining a convenient parametric structure of the function in scope, the learning problem is often defined on an RKHS. Such a machine learning technique is known as the kernel method   \cite{scholkopf2002learning}. 
Let $\mbb{X} \subset \mRd{n}$ contain an infinite number of points. A continuous bivariate function $\kappa: \mbb{X}\times \mbb{X} \rightarrow \mR$ is said to be a \emph{Mercer kernel}, if for any finite number of points $x_1, \dots, x_m \in \mbb{X}$, the $m\times m$ matrix $G_\kappa = [\kappa(x_i, x_j)]$ is positive semidefinite. This implies that $\kappa(x, x)\geq 0$ for all $x\in \mbb{X}$ and $\kappa(x, x') = \kappa(x', x)$ for all $x, x'\in \mbb{X}$. 
The following space:
$$\mc{N}_{\mgk}(\mbb{X}) = \overline{\mr{span}}\{\kappa(x, \cdot): \, x\in \mbb{X}\}$$
endowed with an inner product $\ip{\kappa(x, \cdot)}{\kappa(x', \cdot)} = \kappa(x, x')$, is called the \emph{reproducing kernel Hilbert space}. 

\par We write $\phi: \mbb{X} \rightarrow \mc{N}_{\mgk}(\mbb{X})$, $\phi(x) =: \phi_x = \kappa(x, \cdot)$ as the \emph{canonical feature map}. It satisfies $\ip{\phi_x}{\phi_{x'}} = \kappa(x,x')$ and in fact, for all $f\in \mc{N}_{\mgk}(\mbb{X})$, $\ip{\phi_x}{f} = f(x)$ (which is known as the reproducing property). 
The feature map $\phi: \mbb X\rightarrow \mc{N}_{\mgk}(\mbb{X})$ is continuous. 
Any $f$ in the RKHS $\mc{N}_{\mgk}(\mbb{X})$ is also continuous and bounded, due to the continuity of the canonical feature map $\phi$. Hence $\mc{N}_{\mgk}(\mbb{X})\subset \mCb(\mbb{X})$. Furthermore, provided that $\kappa(x,x)^{1/2}$ is bounded over $x\in \mbb{X}$, the embedding from RKHS to $\mCb(\mbb{X})$ is bounded, since $\|\phi_x\|_{\mc{N}_{\mgk}} \leq \sup_{x\in \mbb{X}} \kappa(x,x)^{1/2}$ and
$$ \|f\|_{\mCb} = \sup_{x\in \mbb{X}} |\ip{\phi_x}{f}| \leq \sup_{x\in \mbb{X}} \|\phi_x\|_{\mc{N}_{\mgk}} \|f\|_{\mc{N}_{\mgk}}. $$

\par We say that the Mercer kernel is \emph{radial} if $\kappa(x, x') = \rho(|x-x'|)$ for some function $\rho: \mR_+ \rightarrow \mR_+$. The following fact, seen in Wendland \cite[Th. 10.35]{wendland2004scattered} and K{\"{o}}hne et al. \cite[Th. 4.1]{kohne2024infty}, confirms the equivalence between the RKHS associated with a radial kernel and the Sobolev--Hilbert space. Let us refer to this kernel as the Sobolev kernel and denote by $\kappa_{\mr{Sob}}$. 
\begin{fact}[Sobolev kernel and Sobolev space]
	If $\hat{\rho}(\omega) = \int_0^\infty \rho(r) \mr{e}^{-\mr{i}\omega r} \dif{r}$ is such that $c_1(1+|\omega|^2)^{-s} \leq |\hat{\rho}(\omega)| \leq c_2(1+|\omega|^2)^{-s}$ for constants $0<c_1\leq c_2$ and $s>n/2$, and if $\mbb{X} \subset \mR^n$ has a Lipschitz boundary, then for the radial kernel $\kappa_{\mr{Sob}}(x,x') = \rho(|x-x'|)$, it holds that $\mc{N}_{\mgk_{\mr{Sob}}}(\mbb{X})\simeq H^s(\mbb{X})$.\footnote{
    A construction of such a radial function $\rho$ was given by Wendland \cite{wendland2004scattered} in the following way. First, let $\rho_l(r) = \max\{1-r, 0\}^l$ ($r\in \mR_+$) for all $l\in \mbb{N}$. By defining operator $I$ on the space of polynomials supported within $[0, 1]$: $(Ig)(r) = \int_r^\infty r'g(r')\dif{r'}$, one can denote $\rho_{n,k} = I^k\rho_{\lfloor n/2 \rfloor + k+1}$ and let $\kappa_{n,k}^{\mr{Wen}}(x, x') = \rho_{n,k}(|x-x'|)$ on $\mbb{X}\subset \mRd{n}$ (for $k\in \mbb{N}$ and $n\geq 1$, where $n\geq 3$ if $k=0$). Then $\mc{N}_{\mgk_{n, k}^{\mr{Wen}}}(\mbb{X}) \simeq H^{\frac{n+1}{2}+k}(\mbb{X})$. 
    Clearly, we may also choose a constant $\sigma>0$ to rescale the kernel, i.e., to use $\tilde\rho(r) = \rho(r/\sigma)$ as the radial function. \label{fn:Wendland}}  
\end{fact}

However, when the system is known to have an equilibrium point at the origin and its stability or stabilization is of interest, Sobolev kernel may not be suitable, as it would lead to the property: $\|\kappa_{\mr{Sob}}(x, \cdot)\|^2 = \kappa_{\mr{Sob}}(x, x) \equiv \rho(0)$ ($\forall x\in \mbb{X}$), i.e., any state is lifted onto the same sphere in the RKHS. 
It is needed, hence, to introduce a new ``equilibrium-aware'' kernel. The following fact was given in the author's earlier work \cite{tang-ye2025koopman}. 
\begin{fact}[Linear--radial kernel and linear--Sobolev space]
    The following function is a Mercer kernel, called the \emph{linear--radial kernel}:
    $$\kappa(x, x') = (x^\top x')\kappa_{\mr{Sob}}(x,x'). $$
    Under the condition that $\mbb{X}\subset \mR^n$ has a Lipschitz boundary, its associated RKHS $\mc{N}_\kappa(\mbb{X})$ coincides with
    $$\textstyle \rg H^s(\mbb{X}) = \BRA{\sum_{k=1}^n e_kg_k: g_1, \cdots, g_n\in H^s(\mbb{X}) }, $$
    where $e_k: x\mapsto x_k$ $(k=1,\cdots,n)$ are the component mappings. $\rg H^s(\mbb{X})$ is a Hilbert space endowed with inner product $\ip{\sum_{k=1}^n e_kg_k}{\sum_{k=1}^n e_kg_k'} = \sum_{k=1}^n \ip{g_k}{g_k'}$ and norm $\|\sum_{k=1}^n e_kg_k\|_{\rg H^s} = \left[ \sum_{k=1}^n\|g_k\|_{H^s}^2 \right]^{1/2}$. 
\end{fact}

It is then direct to verify the property: $\|\phi_x\| = \kappa(x,x)^{1/2} \propto |x|$, which says that any state is lifted to an infinite-dimensional vector whose norm is proportional to its distance from the origin. One can also see that the space $\rg H^s(\mbb{X})$, as an RKHS, contains only functions that are valued $0$ at the origin and have linear terms. 
Thus colloquially, the RKHS of the linear--radial kernel is the \emph{linear--Sobolev space} containing \emph{``locally at least linear''} functions. 
In this paper, the use of $\mc{N}_\kappa(\mbb{X}) \simeq \rg H^s(\mbb{X})$ is two-fold. 
First, given that the system \eqref{eq:controlled.system} has an equilibrium point at the origin, we will define the Koopman operator on the linear--Sobolev type of RKHSs. Second, in the policy evaluation and synthesis, we will define the cost function as quadratic forms of $\phi_x$ under the linear--radial kernel, as \emph{``locally at least quadratic''} functions. 

\par Another instrument that we will use later is the tensor product (as defined in the previous section) of two RKHSs: $\mc{N}_\kappa(\mbb X)$, with canonical feature map $\phi$, and $\mc{N}_{\mgvk}(\mbb U)$, with canonical feature map $\varphi$. 
Due to the reproducing property and the tensor definition, for any $g\in \mc{N}_\kappa(\mbb X)$ and $h\in \mc{N}_\mgvk(\mbb{U})$, we have $\ip{g\otimes h}{\phi_x \otimes \varphi_u} = \ip{g}{\phi_x}\ip{h}{\phi_u} = g(x)h(u)$. Hence, $\mc{N}_\kappa(\mbb X) \otimes \mc{N}_\mgvk(\mbb{U})$ is an RKHS with reproducing kernel $\bar\kappa((x, u), (x',u')) = \kappa(x,x')\varkappa(u,u')$ and canonical feature $\bar\phi_{(x,u)} = \bar\kappa((x, u), (\cdot,\cdot))$.

\subsection{Koopman operator and its learning}\label{subsec:Koopman.learning}
\begin{definition}
	The composition operator, or \emph{Koopman operator}, for an autonomous system $x_{t+1} = f(x_t)$, where $f: \mbb X\rightarrow \mbb X$ is continuous and $\mbb X\subset \mRd{d_x}$, is 
	\begin{equation}\label{eq:KO}
		K_f: \mCb(\mbb X) \rightarrow \mCb(\mbb X), \enspace g\mapsto g\circ f.  
	\end{equation}
    $\mCb(\mbb X)$ is the space of bounded continuous functions on $\mbb X$. 
\end{definition}
The Koopman operator is indeed a well-defined bounded linear operator on $\mCb(\mbb X)$.\footnote{
Since $\mCb(X)$ is a Banach space with norm $\|g\|_{\mCb} = \sup_{x\in X} |g(x)|$. Clearly, for any $g\in \mCb(X)$, we have $\|K_fg\|_{\mCb} = \sup_{x\in X}|g(f(x))| \leq \sup_{y\in X} |g(y)| = \|g\|_{\mCb}$. Hence $\|K_f\|\leq 1$.} 
From a learning point of view, it would more convenient if $K_f$ can be defined directly on some RKHS. Indeed, this can be guaranteed by the regularity of the dynamics $f$. For $K_f$ to be a bounded operator on $\mc{N}_{\kappa_{\mr{Sob}}}(\mbb{X}) \simeq H^s(\mbb{X})$, the key condition given in K{\"{o}}hne et al. \cite{kohne2024infty} is $f\in \mCb^s(\mbb{X}, \mbb{X})$, $s>d_x/2$.\footnote{
Here $\mCb^s(\mbb X)$ for $s\in \mbb{N}$ refers to the space of functions that have bounded derivatives up to order $s$. It is a Banach space with norm $$\textstyle \|g\|_{\mCb^s(\mbb{X})} = \sum_{|\alpha|\leq s}\sup_{x\in \mbb{X}} |g(x)|.$$ For fractional $s = \lfloor s\rfloor + r$, $r\in (0, 1)$, the space $\mCb^s(\mbb{X})$ is understood as an interpolation space between $\mCb^{\lfloor s\rfloor}$ and $\mCb^{\lceil s\rceil}$.} 
In the same spirit as \cite{kohne2024infty}, the following conditions are given for $K_f$ to be a bounded operator on $\mc{N}_\kappa(\mbb{X})\simeq \rg H^s(\mbb{X})$. Here, let us denote
$$ \textstyle \rCb^s(\mbb{X}) = \BRA{\sum_{k=1}^{d_x} e_kg_k: g_k\in \mCb^s(\mbb{X})}, $$
and endow it with norm:
$$\textstyle \norm{\sum_{k=1}^{d_x} e_kg_k}_{\rCb^s(\mbb{X})} = \max_{k=1, \cdots,d_x} \|g_k\|_{\mCb^s(\mbb X)}.$$
\begin{fact}[Koopman operator on the linear--Sobolev space \cite{tang-ye2025koopman}]
	Suppose that $\mbb{X}\subset \mR^{d_x}$ is compact, $f\in \rCb^s(\mbb{X}, \mbb{X})$, $s>d_x/2$ and its Jacobian $\mr{D}f$ is such that $\inf_{x\in X} |\mr{det}\,\mr{D}f(x)| > 0$. Then $K_f \in \mc{L}(\rg H^s(\mbb X), \rg H^s(\mbb X))$. 
\end{fact}

\par Given an independent and identically distributed sample of states $\{x_i\}_{i=1}^m$ and successors $\{y_i = f(x_i)\}_{i=1}^m$, the learning of the Koopman operator becomes the estimation of a linear mapping $\hat{K}_f$ on the RKHS $\mc{N}_{\mgk}(\mbb X)$ such that under the estimated operator, the images of $\{\phi_{y_i}\}_{i=1}^m$, when evaluated on $\{x_j\}_{i=1}^m$, coincides with the actual Koopman operator: 
$$(\hat{K}_f\phi_{y_i})(x_j) = (K_f\phi_{y_i})(x_j) = \kappa(y_i, y_j), \enskip i,j = 1,\dots, m.$$
To this end, we only need to let the estimated Koopman operator be specified by 
$$\textstyle \hat{K}_f\phi_{y_i} = \sum_{j=1}^m \theta_{ij}\phi_{x_j},$$
where the coefficients $\theta_{ij}$ ($i,j=1,\dots,m$), in a matrix $\Theta\in \mRd{m\times m}$, satisfies $\Theta G_{xx} = G_{yy}$. Here the Gramian matrix $G_{xx} = [\kappa(x_i, x_j)]$ and $G_{yy} = [\kappa(y_i, y_j)]$. 

\par Hence, if applied to any observable $g\in \mc{N}_{\mgk}(\mbb X)$ to predict the value of $(K_fg)(x)$ at any $x\in \mbb X$, the estimated Koopman operator acts as if $g$ is first interpolated by $\{\phi_{y_i}\}_{i=1}^m$. Thus,  
\begin{equation}\label{eq:KO.approximation}
	\langle \hat K_f g, \, \phi_x\rangle = \langle K_fSg, \,\phi_x \rangle, 
\end{equation}
where the \emph{kernel interpolation} operator $S: \mc{N}_{\mgk}(\mbb X)\rightarrow \mr{span}\{\phi_{y_i}\}_{i=1}^m$ is defined by $Sg = \sum_{i=1}^m (G^{-1}g_y)_i \phi_{y_i}$, in which $g_y = (g(y_1), \cdots, g(y_m))\in \mRd{m}$. 
The error of such a prediction is bounded as follows K{\"{o}}hne et al. \cite[Th. 3.4]{kohne2024infty}. 
\begin{fact}[Prediction error of the learned Koopman operator]
	If $K_f\in \mc{L}(\mc{N}_{\mgk}(\mbb X), \mc{N}_{\mgk}(\mbb X))$, then it holds for  \eqref{eq:KO.approximation} that: 
	$$\|K_f - \hat{K}_f\|_{\mc{N}_\kappa \rightarrow \mCb} \leq \|\mr{id} - S\|_{\mc{N}_\kappa \rightarrow\mCb} \|K_f\|_{\mc{N}_\kappa \rightarrow \mc{N}_\kappa}.$$
\end{fact}
The first factor on the right-hand side is a uniform interpolation error on the RKHS, which is related to the \emph{fill distance} $\eta_{\mbb X}$ of the sample $\{y_i\}_{i=1}^m$ on $\mbb X$ \cite{wendland2004scattered}:
$$\|\mr{id} - S\|_{\mc{N}_\kappa \rightarrow\mCb} \leq c\eta_{\mbb X}^{s-d_x/2}, \enspace \eta_{\mbb{X}} = \sup_{y\in \mbb X}\min_{i=1,\dots,m} \|y-y_i\|,$$
where $c$ is a constant dependent only on $s$, $d_x$, and $\mbb X$. 
Given the scaling law of fill distance $\eta_{\mbb X}\propto m^{-1/d_x}$, we can write 
$$\|K_f - \hat{K}_f\|_{\mc{N}_\kappa \rightarrow \mCb} \lesssim m^{-\bra{\frac{s}{d_x}-\frac{1}{2}}}.$$

\section{Koopman--Nemytskii Operator}\label{sec:Definition}
Now we consider an (unknown) nonlinear system in the form of \eqref{eq:controlled.system}. We make the following standing assumptions for the well-definedness of the Koopman operator under all feedback laws in consideration. 
\begin{assumption}\label{assum:1}
	For all policies $u\in \mbb U$, $f_u\in \rCb^{s+1}(\mbb X, \mbb{X})$, with $s> d_x/2$, $\inf_{x\in \mbb X, u\in \mbb U} |\mr{det}\,\mr{D}f_u(x)| > 0$, and $\mbb{X} \subset \mR^{d_x}$ is a compact set with Lipschitz boundary. 
\end{assumption} 
Thus, by using the linear--radial kernel $\kappa$ such that $\rg H^{s+1}(\mbb X) \simeq \mc{N}_\mgk(\mbb{X})$, we have 
$$K_{f_u}\in \mc{L}(\rg H^{s+1}(\mbb X), \rg H^{s+1}(\mbb X)) \simeq \mc{L}(\mc{N}_\mgk(\mbb{X}), \mc{N}_\mgk(\mbb{X})). $$
Now that the Koopman operator $K_{f_u}$ describes the closed-loop dynamics under a given policy $u$, we examine the dependence of $K_{f_u}$ on $u\in \mbb U$. Out of this examination, the concept of Koopman--Nemytskii operator will arise. 
\begin{remark}[Special case of state-independent policies]
    The discussions in the sequel aims to derive the operator model \eqref{eq:conceptual.policy-state.viewpoint} for policies $u$ in a prespecified class $\mbb{U}$. This construction contains the special case of state-irrelevant constant-valued policies, i.e., $\mbb U=\{x\mapsto a: a\in \mbb A\}$. This special case reduces to modeling the open-loop system under different input \emph{values} instead of \emph{policies}, i.e., will result in the ``input--state operator model'' \eqref{eq:conceptual.input-state.viewpoint}. 
\end{remark}

\subsection{Dependence of the Koopman operator on the policy}
Clearly, the closed-loop dynamics $f_u = f(\cdot, u(\cdot))$ depends on the policy $u$. If the function $f$ is sufficiently smooth, then the dependence of $f_u$ on $u$ is naturally continuous. 
\begin{definition}
	Suppose that $f\in \rCb^s(\mbb X \times \mbb A, \mbb X)$ for some $s\in \mR_+$. The substitution operator, or \emph{Nemytskii operator}, of the controlled system \eqref{eq:controlled.system} refers to:
	\begin{equation}
		N_f: \, \rCb^s(\mbb X, \mbb A) \rightarrow \rCb^s(\mbb X, \mbb X), \, u \mapsto f(\cdot, u(\cdot)) = f_u,
	\end{equation}
	namely the mapping from the set of feedback laws to corresponding closed-loop dynamics.
\end{definition}
The definition requires that $f_u\in \rCb^s(\mbb{X}, \mbb{X})$ whenever $u\in \rCb^s(\mbb X, \mbb A)$. This is indeed true, briefly proved below. 
\begin{proof}
    By the condition that $f\in \rCb^s(\mbb X\times \mbb A, \mbb X)$ we can write 
    $$\textstyle f(x, a) = \sum_{k=1}^{d_x} x_kp_k(x, a) + \sum_{l=1}^{d_u} a_lq_l(x, a)$$
    with all $p_k, q_l\in \mCb^s(\mbb X \times \mbb A, \mbb X)$. Thus 
    $$\textstyle f(x, u(x)) = \sum_{k=1}^{d_x} x_kp_k(x, u(x)) + \sum_{l=1}^{d_u} u_l(x)q_l(x, u(x)). $$
    Since $u_l \in \rCb^s(\mbb{X})$, we have $u_l(x) = \sum_{k=1}^{d_x} x_kw_{lk}(x)$ with $w_{lk}\in \mCb^s(\mbb{X})$. Thus 
    $$\textstyle f(x, u(x))=\sum_{k=1}^{d_x} x_k [ p_k(x, u(x)) + \sum_{l=1}^{d_u} w_{lk}(x)q_l(x, u(x))] ,$$
    where $p_k(\cdot, u(\cdot)) + \sum_{l=1}^{d_u} u_{lk} q_l(\cdot, u(\cdot)) \in \mCb^s(\mbb{X}, \mbb{X})$. Therefore $f(\cdot, u(\cdot))=f_u\in \rCb^s(\mbb{X}, \mbb{X})$. 
\end{proof}
\begin{proposition}
	At any fixed $u\in\rCb^s(\mbb X, \mbb A)$, the Nemytskii operator $N_f$ is continuous, if $f\in \rCb^{s+1}(\mbb X\times \mbb A, \mbb X)$. 
\end{proposition}
\begin{proof}
	Without loss of generality, only consider $d_x = d_u =1$. The proposition is proved by induction. When $s=0$, suppose that $f\in \rCb^1(\mbb{X}, \mbb{X})$. Then in a similar way to the previous proof, let $f(x) = xp(x, a)+aq(x, a)$. Consider $u(x)=xw(x)$ and $(u+\tilde u)(x) = x(w(x)+\tilde w(x))$. Then $f(x, u(x))= x[p(x, xw(x))+w(x)q(x,xw(x))]$, which implies that  
	$$\textstyle  \|f_{u+\tilde{u}} - f_u\|_{\rCb^0} = \|p_{w+\tilde{w}} - p_w + (w+\tilde{w})q_{w+\tilde{w}} - wq_w \|_{\mCb^0}, $$
    where $p_w(x) = p(x, xw(x))$ and $q_w(x) = q(x, xw(x))$. Now that $p, q\in \mCb^1(\mbb{X})$ by Assumption \ref{assum:1}, obviously, by 
    $$\textstyle  \|p_{w+\tilde{w}} - p_w\|_{\mCb^0} \leq \|p\|_{\mCb^1}\|\tilde{w}\|_{\mCb^0}$$
    and the analogous inequality for $q$, we obtain $\|N_f(u+\tilde{u}) - N_fu\|_{\rCb^0} \leq \|f\|_{\rCb^1} \|\tilde{u}\|_{\rCb^0}$. The proposition holds for $s=0$. 
    \par Suppose that the proposition holds for all $0\leq r\leq s-1$:
	$$\|N_f(u+\tilde{u}) - N_fu\|_{\rCb^r} \leq \|f\|_{\rCb^{r+1}} \|\tilde{u}\|_{\rCb^r}. $$
	For $(N_f(u+\tilde u) - N_f u)(x) = x[p_{w+\tilde{w}}(x)+ (w+\tilde w)q_{w+\tilde{w}}(x)-p_w(x)-wq_w(x)]$, consider a partial derivative $\mr{D}^s$ acted on the bracketed term after the $x$-term on the right-hand side. 
    By chain rule, this partial derivative comprises of differences in counterpart terms resulting from $w+\tilde w$ and $w$, each being a product of partial derivatives of $p$, $q$, and $\mr D^r w(x)$ ($r\leq s$). All these differences are bounded by constant multiples of $\|\tilde{w}\|_{\mCb^s} = \|\tilde{u}\|_{\rCb^s}$, in which the coefficient involves $\|f\|_{\rCb^{s+1}}$. Therefore, the proposition then holds true for $s$. 
\end{proof}

\par Then, we consider the dependence of the Koopman operator $K_{f_u}$ on the dynamics $f_u$. 
\begin{definition}\label{def:M}
	Let $s\in \mbb{N}$. The \emph{``Koopmanizing'' operator} is  
	$$M: \mc{F} \rightarrow \mc{L}(\rg{H}^{s+1}(\mbb X), \rg{H}^s(\mbb X)), \enskip f\mapsto K_f, $$
	where $K_f$ is considered as a mapping from $\rg{H}^{s+1}(\mbb X)$ to $\rg{H}^s(\mbb X)$, and $\mc{F} \subset \rCb^{s+1}(\mbb X, \mbb X)$ is a family of functions that guarantees $\inf_{x\in \mbb X}|\mr{det}\, \mr{D}f(x)|>0$ for all $f\in \mc{F}$. 
\end{definition}
\begin{proposition}
	Under the conditions given in Definition \ref{def:M}, $M$ is a continuous operator. 
\end{proposition}
\begin{proof}
    Induction is used to show the conclusion. First consider $s=0$. At any fixed $f\in \mc{F}$, we check $\|M(f+\tilde{f}) - Mf\|$ due to a small variation $\tilde{f}$ in $\rCb^1$ such that $f+\tilde{f}\in \mc{F}$: 
	$$\textstyle \|K_{f+\tilde{f}} - K_f\| = \sup_{\|g\|_{\rg{H}^1} \leq 1} \|g(f(\cdot) + \tilde{f}(\cdot)) - g(f(\cdot)) \|_{\rg H^0}.$$
	Without loss of generality, consider the case with $d_x=1$ and $d_u=0$. Then $g(x)=x\gamma(x)$ and $f(x)=xp(x)$. Thus, 
    $$\textstyle  g(f(x)+\tilde{f}(x)) - g(f(x)) = x [\bar\gamma(x, p(x)+\tilde{p}(x)) -  \bar\gamma(x, p(x))] $$
    where $\bar\gamma (x,p) := p\gamma(xp)$. Letting $\bar\gamma_p(x) = \bar\gamma(x,p(x))$, we get
    $$\textstyle  \|M(f+\tilde{f}) - Mf\|^2 = \int_{\mbb X} \bra{ \bar\gamma_{p+\tilde p} - \bar\gamma_p}^2 \dif{x}. $$
    Since $\gamma$ (and hence $\bar\gamma$) is in $\rg H^1(\mbb X)$, $\mr{D}_p \bar\gamma(x, p) = \gamma(xp)+xp\mr{D}\gamma(xp)$, we have the integrand upper-bounded by a constant multiple of $\|\tilde p\|_{\mCb^0}^2 \int_{\mbb{X}} \bra{\gamma(xp(x))^2 + \mr{D}\gamma(xp(x))^2 } \dif{x}$. With the assumption that $|\mr{det}\, \mr{D}f(x)|\geq \epsilon>0$ for some $\epsilon>0$, we have the integration further upper bounded, through change of variables $y=f(x)$, by $\int_{\mbb X} \bra{ \gamma(y)^2 + \mr{D}\gamma(y)^2} \dif{y}$, namely $\|\gamma\|_{H^1}^2 = \|g\|_{\rg H^1}^2$. 
	Therefore, upon $\|g\|_{\rg{H}^1}\leq 1$, we obtain $\|M(f+\tilde{f}) - Mf\| \lesssim \|\tilde{f}\|_{\rg C_0}$. 
    \par Suppose that for all $0\leq r\leq s-1$:
	$$\textstyle  \|M(f+\tilde{f}) - Mf\|_{\rg{H}^{r+1} \rightarrow \rg{H}^r} \lesssim \|\tilde{f}\|_{\rg C^r}.$$
	Then in the case of $s$, for $g(f(x)) = xp(x)\gamma(xp(x)) = x\bar\gamma(x,p(x))$, the action of $\mr{D}^s$ on it results in terms that involve the $r$-th partial derivatives of $\bar\gamma$ (hence $\gamma$) for $0\leq r\leq s$. When taking the difference between $f+\tilde{f}$ and $f$, each of the comprising terms, upon squared integration, will be bounded by a constant multiple of $\|\gamma\|_{H^{s+1}}^2$ (and hence $\|g\|_{\rg H^{s+1}}^2$) multiplied by $\|\tilde p\|_{C^s(\mbb{X})}$. Therefore, upon $\|g\|_{\rg{H}^{s+1}} \leq 1$, 
	$$\textstyle  \|g(f(\cdot) + \tilde{f}(\cdot)) - g(f(\cdot)) \|_{\rg{H}^{s}} \lesssim \|\tilde{f}\|_{\rg C^s}.$$
    The proof is then completed. 
\end{proof}
 
\par Composing the two operators, the operator $MN_f: u\mapsto K_{f_u}$
becomes a nonlinear but continuous operator from $\rCb^{s+1}(\mbb X, \mbb A)$ to $\mc{L}(\rg{H}^{s+1}(\mbb X), \rg{H}^s(\mbb X))$. 
The Sobolev index of the codomain is $1$ lesser than the domain. This is inevitable since the effect of any variation in policy on the closed-loop dynamics is embodied on the observable $g$ through the derivatives of $g$. 
As the norm in $\rg{H}^{s+1}$ is stronger than the norm in $\rg{H}^s$, it becomes impossible to deem $MN_f$ as a continuous operator to $\rg{H}^{s+1}$ or an operator from $\rg{H}^s$. This issue does not arise for autonomous systems (e.g., in \cite{kohne2024infty} or \cite{tang-ye2025koopman}). 
Due to this reason, next, instead of considering the continuous dependence of $K_{f_u}$ on $u$, we focus on its adjoint operator $K_{f_u}^\ast \in \mc{L}(\rg{H}^s(\mbb X), \rg{H}^{s+1}(\mbb X))$.

\subsection{Definition of the Koopman--Nemytskii operator}\label{subsec:KN}
\par By the adjoint operator $K_{f_u}^\ast$, we refer to the one such that
$${\ip{K_{f_u}^\ast h}{h'}}_{\rg{H}^{s+1}} = {\ip{h}{K_{f_u}h'}}_{\rg{H}^s}, \forall h\in \rg{H}^s(\mbb X), h'\in \rg{H}^{s+1}(\mbb X).$$
The adjoint operator of the Koopman operator can be called as \emph{Perron--Frobenius operator}.  

\begin{proposition}
	The operator defined by  
	$$T_0: \mbb{U} \subset \rCb^{s+1}(\mbb{X}, \mbb{A}) \rightarrow \mc{L}(\rg{H}^s(\mbb X), \rg{H}^s(\mbb X)), \enskip u\mapsto K_{f_u}^\ast$$
	is continuous under Assumption \ref{assum:1}, and has the property:
	\begin{equation}\label{eq:kernel.correspondence}
		(T_0u)\phi_x = \phi_{f_u(x)}, \enskip \forall u\in \mbb U, x\in \mbb X, 
	\end{equation}
	where $\phi$ is the canonical map of the linear--radial kernel, the reproducing kernel of $\rg H^{s+1}(\mbb X)$.
\end{proposition}
\begin{proof}
	Since $MN_f$ is continuous, at any $u\in \mbb U$, as $u'\rightarrow u$, $\|K_{f_u} - K_{f_{u'}}\|\rightarrow 0$, implying that $\|T_0u' - T_0u\|_{\rg{H}^{s+1}} = \|K_{f_u} - K_{f_{u'}}\|\rightarrow 0$. Since the $\rg{H}^s$-norm is weaker than the $\rg{H}^{s+1}$-norm, $\|T_0(u'-u)\|_{\rg{H}^s} \rightarrow 0$. Hence, $T_0$ is continuous. 
    To verify \eqref{eq:kernel.correspondence}, take any $g\in \rg{H}^s(\mbb X)$, we verify $\langle K_{f_u}^\ast\phi_x, g \rangle = \ip{\phi_x}{K_{f_u}g} = (K_{f_u}g)(x) =  g(f_u(x))$, which implies $K_{f_u}^\ast\phi_x = \phi_{f_u(x)}.$
\end{proof}

\par To resolve the nonlinearity, we ``lift'' the policy space $\mbb U$ into a new RKHS. 
That is, we define a Mercer kernel $\mgvk$, which assigns a $\mgvk(u_1,u_2)\in \mR$ to every pair of policies $(u_1, u_2)\in \mbb U\times \mbb U$. We denote the canonical feature map of this kernel as $\varphi$, i.e., $\varphi_u = \mgvk(u, \cdot)$ ($\forall u\in \mbb U$). Hence, an RKHS $\mc{N}_{\mgvk}(\mbb U)$ is defined.\footnote{
    The creation of such a kernel is always possible. Since $\mbb U\subset \rCb^{s+1}$, we can assign any injective mapping $\phi$ from $\mbb U$ to a Hilbert space, and let $\mgvk(u, u') = \ip{\phi(u)}{\phi(u')}$. 
	For example, if $\mbb X$ is a bounded region, $\rCb^{s+1}(\mbb X)$ is a subspace of the linear--Sobolev space $\rg{H}^{s+1}(\mbb X)$, and thus we may define a distance on $\mbb U$ as $d(u,u') := \left[\sum_{|\beta|\leq s+1} \int_{\mbb X} \|\mr{D}^\beta (u-u')\|^2\right]^{1/2}$, and use a Gaussian kernel: $\mgvk(u,u') = \mr{e}^{-d(u,u')^2/\sigma^2}$ for some constant $\sigma>0$.
	}
When the family of feedback laws is sufficiently smoothly parameterized: $u = u(\cdot|v)$, with parameters $v\in \mbb{V} \subset \mR^{d_v}$ continuously mapped to $u(\cdot|v)\in \rCb^{s+1}(\mbb{X}, \mbb{A})$, then it suffices to define the kernel simply on $\mbb{V}$. 
In general, when $U$ is a compact metric space, a \emph{universal kernel} can always be created, which means that the associated RKHS is dense in $C(\mbb U)$, the space of continuous functionals of policies (see, e.g., Steinwart and Christmann \cite{steinwart2008support}). That is, any $g\in C(\mbb U)$ can be approximated by an $h_\epsilon$ belonging to the RKHS $\mc{N}_{\mgvk}(\mbb U)$ such that $\|h_\epsilon - g\|_{C(\mbb U)} < \epsilon$ to an arbitrary precision $\epsilon$. 

\begin{assumption}\label{assum:2}
	$\mbb U$ is compactly contained in $\rCb^{s+1}(\mbb X, \mbb A)$, and $\mgvk$ is a universal kernel on $\mbb U$, with $\varkappa(u,u)\leq 1$ ($\forall u\in \mbb U$). 
\end{assumption}
\begin{theorem}
	Under Assumptions \ref{assum:1} and \ref{assum:2}, there exists a linear bounded operator
	\begin{equation}\label{eq:KN.operator.pre}
		T_1: \mc{N}_{\mgvk}(\mbb U)\rightarrow \mc{L}(\rg{H}^s(\mbb X), \rg{H}^s(\mbb X)), \enspace \varphi_u \mapsto K_{f_u}^\ast. 
	\end{equation}
\end{theorem}
\begin{proof}
	\par Due to the previous proposition, $T_0$ is a continuous mapping from $\mbb U$ to $\mc{L}(\rg{H}^s(\mbb X), \rg{H}^s(\mbb X))$. Hence, for any linear bounded functional on $\mc{L}(\rg{H}^s(\mbb X), \rg{H}^s(\mbb X))$, i.e., $L\in \mc{L}(\rg{H}^s(\mbb X), \rg{H}^s(\mbb X))^\ast \simeq \mc{L}(\rg{H}^s(\mbb X), \rg{H}^s(\mbb X))$, $LT_0$ is a continuous mapping from $U$ to $\mR$. 
	Given that $\mgvk$ is universal, $LT_0 \in C(\mbb U)$ can be approximated by a corresponding member of $\mc{N}_{\mgvk}(\mbb U)$, which is a linear functional acting on $\varphi_u$. 
    That is, $\forall \epsilon>0$, $\exists v_L^\epsilon \in \mc{N}_{\mgvk}(\mbb U)^\ast \simeq \mc{N}_{\mgvk}(\mbb U)$, such that 
	$$|LT_0u - \ip{v_L^\epsilon}{\varphi_u}| < \epsilon, \enskip \forall u\in \mbb U.$$
	\par Choose a sequence $\{\epsilon_j\}_{j\in \mbb{N}} \downarrow 0$ (e.g., $\epsilon_j = 1/j$), and examine $v_L^{1/j}$. When $j$ is large enough, for any $k\in\mbb{N}$, we have $\left|\ip{v_L^{1/j} - v_L^{1/(j+k)}}{\varphi_u} \right| < 1/j + 1/(j+k) < 2/j$ for all $u\in \mbb U$. 
    Hence, for any $h\in \mc{N}_{\mgvk}(\mbb U)$ with an RKHS norm not exceeding $1$, it holds that $\left|\ip{v_L^{1/j} - v_L^{1/(j+k)}}{h}\right| < 2/j \rightarrow 0$ ($j\rightarrow \infty$). Therefore, the sequence $\{v_L^{\epsilon}\}$ with $\epsilon\downarrow 0$ weakly converges in $\mc{N}_{\mgvk}(\mbb{U})$. Obviously, the weak limit is unique, which we denote by $v_L$. It satisfies: 
	$$LMN_fu = \ip{v_L}{\varphi_u}, \enskip \forall u\in \mbb U.$$
	\par We note that $v_L$ depends linearly on $L$. Thus, the mapping 
	$$V: \mc{L}(\mc{H}^s(\mbb X), \mc{H}^s(\mbb X))^\ast \rightarrow \mc{N}_{\mgvk}(\mbb U)^\ast, \enskip L\mapsto v_L$$
	is linear. Its adjoint $V^\ast: \mc{N}_{\mgvk}(\mbb U) \rightarrow \mc{L}(\rg{H}^s(\mbb X), \rg{H}^s(\mbb X))$ is the desired operator mapping each $\varphi_u$ to $T_0u$. 
\end{proof}

\par With the above construction, we have an operator $T_1\in \mc{L}(\mc{N}_{\mgvk}(\mbb U), \mc{L}(\rg{H}^s(\mbb X), \rg{H}^s(\mbb X)))$, from a RKHS to an operator space. Naturally, the operator space that $T_1$ resides in is equivalent to $\mc{L}(\rg{H}^s(\mbb X) \otimes \mc{N}_{\mgvk}(\mbb U), \rg{H}^s(\mbb X))$, where we simply bring the second argument (namely the state-dependent function) into the beginning position. 
\begin{definition}
	Under Assumptions \ref{assum:1} and \ref{assum:2}, the \emph{Koopman--Nemytskii operator} is defined as 
	a linear bounded operator 
	\begin{equation}\label{eq:KN}
		T: \rg{H}^s(\mbb X) \otimes \mc{N}_{\mgvk}(\mbb U) \rightarrow \rg{H}^s(\mbb X), \enspace \phi_x\otimes \varphi_u \mapsto \phi_{f_u(x)}.
	\end{equation}
\end{definition}
Here the domain is the tensor product of $\mc{H}^s(\mbb X) \simeq \mc{N}_{\kappa}(\mbb X)$ and $\mc{N}_{\mgvk}(\mbb U)$, which is an RKHS with a reproducing kernel $\bar{\mgk}$ specified by $\bar\mgk((x, u), (x',u')) = \kappa(x,x')\varkappa(u, u')$. 
In other words, $\mc{N}_{\bar{\mgk}}(\mbb X \times \mbb U) = \mc{N}_{\kappa}(\mbb X) \otimes \mc{N}_{\mgvk}(\mbb U)$. Denoting its canonical map as $\bar{\phi}$, we write $\bar{\phi}_{(x, u)} = \phi_x \otimes \varphi_u$. 
Hence, 
$$\begin{aligned}
    & T\in \mc{L}(\mc{N}_{\bar{\mgk}}(\mbb X \times \mbb U), \, \mc{N}_{\mgk}(\mbb X)), \\
    & T\bar{\phi}_{(x, u)} = \phi_{f_u(x)}, \enspace \forall x\in\mbb X, u\in\mbb U.
\end{aligned}$$ 

\begin{remark}[Stochastic interpretation]
    The interpretation of the Koopman--Nemytskii operator is intuitive. Given a state $x\in \mbb X$ and a policy $u\in \mbb U$, represented by their canonical features, the Koopman--Nemytskii operator returns the canonical feature of the succeeding state $\phi_{f_x(u)}$. 
    Given a ``stochastic mixture of states'' $\sum_i p_i\phi_{x_i}$ and $\sum_j q_j \varphi_{u_j}$ as a ``stochastic mixture of policies'' (which may not be necessarily normalized to $\sum_i p_i = \sum_j q_j = 1$), the Koopman-Nemytskii operator returns a corresponding ``stochastic mixture of updated states'' $T\left(\sum_i p_i\phi_{x_i}, \sum_j q_j \varphi_{u_j}\right) = \sum_i \sum_j p_iq_j \phi_{f_{u_j}(x_i)}. $ 
    Hence, if the canonical feature of the initial state $x$ can be approximately seen as $\phi_x = \sum_i p_i\phi_{x_i}$, and the policy $u$ is considered as a combination of sampled policies' features: $\varphi_u = \sum_j q_j\varphi_{u_j}$, then the prediction follows the ``mixture'' formula.
\end{remark}

\begin{remark}[Regularity of the system]
\label{rem:regularity}
    The construction of the Koopman--Nemytskii operator requires sufficient smoothness of the dynamics and non-degeneracy of Jacobian. 
    The former condition is posed due to the need for the equivalence between the Sobolev--Hilbert space and an RKHS used in learning \cite{kohne2024infty}. 
    This can be well satisfied by systems whose governing equations arise from physical laws that are smooth. 
    The second condition is also mild --- if the physical dynamics is continuous-time and discretized with a short sampling time, the non-degeneracy of Jacobian is guaranteed by the existence and regularity theory of ordinary differential equations. 
\end{remark}

\begin{remark}[Black-box nature and hybrid modeling] 
\label{rem:hybrid}
    The ``lifting'' of a nonlinear system into an operator model, by itself, overlooks the physical meanings of the underlying dynamics. Hence, the interpretation of the Koopman--Nemytskii operator is only ``empirical''. 
    Users who are concerned with the physical interpretability can devise a hybrid modeling strategy. Possible approaches include (i) collecting simulation data from a low-fidelity first-principles model, and then training the operator on a mixture of low-fidelity simulations and high-fidelity plant data, and (ii) training a reference operator on the simulation data from a low-fidelity first-principles model, and then regularizing the learned operator near the reference when learning from high-fidelity data. 
\end{remark}

\begin{remark}[Parameterization of policy space and need for prior knowledge]
    While in principle, the Koopman--Nemytskii operator can be defined for a wide policy class $\mbb{U}$, or even the entire $\rCb^{s+1}(\mbb X, \mbb A)$, practically, if such an operator is learned from data, the policy class $\mbb U$ must be (low-dimensionally) parameterized. 
    Essentially, this is due to the fact that the policy's effect on the closed-loop dynamics is characterized in a ``black-box'' manner from a Koopman operator learning perspective. In other words, one is offered with no oracle other than the regularity of the defining spaces. 
    Hence, if there is prior information on the forms of satisfactorily performing controllers (based on user's common sense of the system's governing physics), then the policy space $\mbb U$ can be chosen to contain such forms, and the policy kernel $\varkappa$ can be defined on the controller parameters. 
\end{remark}

\subsection{Koopman--Nemytskii operators in continuous time}
\par The approach above is proposed for discrete-time systems mainly due to its formal and computational simplicity. Here we comment on the case of a continuous-time system: 
$$\dif{x_t}/\dif{t} = f(x_t, u(x_t)) =: f_u(x_t).$$
First, let 
$$L: \mc{D}\times \mbb{U} \to \rg H^{s}(\mbb X), \enspace (g, u) \mapsto \mr{D} g\cdot f_u$$
where the domain for the $g$ argument, $\mc{D}$, contains at least a dense subset $\rg C^\infty(\mbb{X})$ of $\rg H^s(\mbb{X})$. Such an operator, under any $u\in \mbb{U}$ is the infinitesimal generator of the Koopman semigroup $K_u^t: g \mapsto g\circ S_u^t$ on $\rg H^s(\mbb{X})$, in which $S_u^t$ represents the flow under $f_u$. 
This infinitesimal generator is closed and densely defined, linear with respect to the $g$ argument (state-dependent function argument), despite an unbounded operator. 
Then, due to the linearity in $g$, by the adjoint relation 
$$ \ip{h}{L(g, u)} = \ip{T_0(h, u)}{g}, $$
for $g\in \mc{D}\subset \rg H^s(\mbb{X}), h\in \mc{D}' \subset \rg H^s(\mbb{X}), u\in \mbb{U}$, a mapping $T_0$ is defined on a dense subspace of $\rg H^s(\mbb{X})$ on which the right-hand side of the above relation specifies a bounded linear mapping on $g$. 
Clearly, $T_0$ is continuous in $u$. Subsequently, following the same procedure as in the previous subsection, we ``lift'' it to an unbounded operator:
\begin{equation}\label{eq:KN.continuous-time}
    T: \mc{D}'\otimes \mbb{U} \subset \mc{N}_\kappa(\mbb X) \otimes \mbb{U} \to \mc{N}_\kappa(\mbb X). 
\end{equation}
Such defined $T$ is said to be the \emph{continuous-time Koopman--Nemytskii operator/generator}.\footnote{While $\phi_x$ may not be in $\mc{D}'$, it can be verified that $\frac{1}{\tau}\int_0^\tau \phi_{S_u^t(x)} \dif{t} \in \mc{D}'$ (where $\tau>0$ is finite). In fact, it can be verified that
$$\textstyle T\bra{\frac{1}{\tau}\int_0^\tau \phi_{S_u^t(x)} \dif{t} , \, \varphi_u} = \frac{1}{\tau}\bra{\phi_{S_u^\tau(x)} - \phi_{x}}, \enspace x\in \mbb X, u\in \mbb U. $$
This relation can be used for learning in a similar way to the discrete-time case as discussed in the next section.}

\par In continuous time, many systems can be expressed in an \emph{input-affine} form (instead of being generally nonlinear):
$$\textstyle \dif{x_t}/\dif{t} = f_0(x_t) + \sum_{j=1}^{d_a} a_{j, t} f_j(x_t), $$
and in the case that $f_0\equiv 0$, the system said to be \emph{holonomic}. 
In such scenarios, $T_0$ and thus $T$ are known to be affinely dependent on $u$ (and linearly, in the holonomic case).
If the policy class in scope $\mbb U$ is affinely (or linearly) parameterized: $u(x|v) = U(x) v = u_0(x) + v_1u_1(x) + \cdots + v_{d_v}u_{d_v}(x)$ for a dictionary of ``policy basis'' $u_1(x), \cdots, u_{d_v}(x)$, then it suffices to directly use $v\in \mR^{d_v}$ as $\varphi_u$ in the continuous-time Koopman--Nemytskii operator.

\section{Learning of Koopman--Nemytskii Operator}\label{sec:Learning}
\par In this section, we consider the data-based approximation of the Koopman-Nemytskii operator $T$ as defined in \eqref{eq:KN}. 

\subsection{Problem setting}\label{subsec:kernel.RRR}
\par Suppose that we have an available dataset $\{(x_i, u_i, y_i): \, y_i = f(x_i, u_i(x_i)), \, i=1,\dots,m\}$. The state-policy combinations $(x_i, u_i)$ are assumed to be generated independently from a joint distribution $\mP$. 
With the linear--radial kernel $\mgk$ for $\rg{H}^s(\mbb X)$, universal kernel $\mgvk$ on $\mbb U$, and thus $\bar{\mgk} = \mgk \otimes \mgvk$ defined, we recall that the true Koopman--Nemytskii operator satisfies: $ T\bar{\phi}_{(x_i,u_i)} = \phi_{f_u(x)}$, $\forall x\in \mbb X, \, u\in \mbb U$. 
Hence, with an \emph{empirical risk minimization} formulation, the learning procedure determines $\hat{T}$ by minimizing the following loss:
$$\textstyle \hat{\ell}_\beta(\hat{T}) := \frac{1}{m}\sum_{i=1}^m \norm{\phi_{y_i} - \hat{T}\bar{\phi}_{(x_i, u_i)} }_{\mc{N}_{\mgk}}^2 + \beta\|\hat{T}\|_{\ms{HS}}^2.$$
Here $\hat{T}$ is searched among Hilbert--Schmidt operators\footnote{An operator $A$ is Hilbert--Schmidt on a separable Hilbert space $\mc{H}$, denoted by $A\in \ms{HS}(\mc{H})$, if $AA^*$ belongs to the trace class. That is, for any orthonormal basis of this Hilbert space $\{e_j\}_{j=1}^\infty$, we have $AA^* = \sum_j \alpha_j e_j\times e_j$ with $\sum_j \alpha_j < \infty$. The Hilbert--Schmidt norm of $A$ is defined as $\|A\|_{\ms{HS}} = (\sum_j \alpha_j)^{1/2}$, independent of the choice of orthonormal basis.} 
from $\mc{N}_{\bar\mgk}(\mbb X\times \mbb U)$ to $\mc{N}_\kappa(\mbb{X})$ with a rank restriction: $\operatorname{rank}\hat{T}\leq r$.
The regularization coefficient $\beta>0$ penalizes the learned ``model complexity'', namely the squared Hilbert--Schmidt norm. 
Such a formulation is called \emph{reduced rank regression} \cite{kostic2022learning}. 

\par The space of Hilbert-Schmidt operators is a Hilbert space itself. Therefore, according to the representer theorem \cite{scholkopf2001generalized}, the minimizer is necessarily a finite-rank operator:
\begin{equation}\label{eq:finite.rank}
	\textstyle \hat{T} = \sum_{i,j=1}^m \theta_{ij} \phi_{y_i}\times \bar{\phi}_{(x_j, u_j)}, 
\end{equation}
where $\Theta = [\theta_{ij}] \in \mRd{m\times m}$ is to be minimized in a finite-dimensional convex optimization problem. 
In Kostic et al. \cite{kostic2022learning}, it was shown that the reduced rank regressor guarantees a statistical bound on the generalized $L^2$-error, i.e., the mean-squared error when predicting the succeeding state feature based on state-policy pairs:
$$\textstyle \ell_0(\hat{T}) = \mbb{E}_{y=f_u(x), \, (x, u)\sim\mbb{P}}\Bra{ \norm{\phi_{y} - \hat{T}\bar\phi_{(x,u)}}_{\mc{N}_\kappa}^2 }. $$

\subsection{Kernel EDMD and its generalization error}\label{subsec:kernel.DMD}
\par Along the lines of Korda and Mezi{\'{c}} \cite{korda2018convergence} and K{\"{o}}hne et al. \cite{kohne2024infty}, when $\hat{T}$ is estimated without using regularization or rank constraint, i.e., by minimizing $\hat \ell_0(\hat{T})$, the resulting coefficient matrix $\Theta$ is uniquely determined by letting \eqref{eq:finite.rank} satisfy:
$$\textstyle \phi_{y_j'} = \hat{T}\bar{\phi}_{(x_{j'}, u_{j'})} = \sum_{i,j=1}^m \theta_{ij}G_{xu, jj'} \phi_{y_i}, \enskip j'=1,\dots,m.$$
Hence, $\Theta = G_{xu}^{-1}$. This approach is called \emph{kernel extended dynamic mode decomposition} (kernel EDMD).\footnote{
Since the kernel matrix $G_{xu}$ can be ill-conditioned, it is more practical to keep the regularization parameter $\beta$. As shown in \cite{bold2025kernel} based on \cite{wendland2005approximate}, the interpolation error depends on both the fill distance $\eta$ and the regularization parameter $\beta$; by tacitly assuming that $\beta$ can be fine-tuned in concert with $\eta$, the dependence on $\beta$ can be removed without changing our conclusions. 
}
Let $S$ stand for the projection from the RKHS $\mc{N}_{\mgk}(\mbb X)$ to its subspace $\mr{span}\{\phi_{y_i}\}_{i=1}^m$. Analogous to the discussions in \S\ref{subsec:Koopman.learning}, the estimated Koopman--Nemytskii operator satisfies, for any $g\in \mc{N}_\kappa(\mbb X)$ and any $(x,u)\in \mbb X\times \mbb U$:
$$ \langle g, \, \hat{T}\bar\phi_{(x,u)} \rangle = \langle Sg, \, T\bar\phi_{(x,u)} \rangle, $$

\par When the sample points cover $\mbb X\times \mbb U$ well, the error of applying the estimated Koopman--Nemytskii operator to the prediction of succeeding state is \emph{uniformly proportional to $|x|$} on $\mbb X\times \mbb U$, instead of being a mean-squared one. This will be proved in Theorem \ref{th:EDMD.generalization}. 
To guarantee such a dense coverage by sample points, it is technically desirable to ensure the finite dimensionality of the policy space $\mbb U$ by a parametric structure. 
\begin{assumption}\label{assum:3}
	$\mbb X$ satisfies the interior cone condition.\footnote{
    We refer to the following set as a cone in $\mRd{d}$: 
    $$\mbb K(x, \xi, \theta, r) = \{x+\lambda y: \|y\| = 1, \ip{y}{\xi} \geq \cos\theta, \lambda \in [0, r]\}, $$ 
    where the vertex $x\in \mRd{d}$, direction $\xi\in \mRd{d}$ a unit vector, angle $\theta\in (0, \pi/2)$ and radius $r>0$. The set $\mbb X\subset \mRd{d}$ is said to satisfy the interior cone condition if there exists a $\theta\in(0,\pi/2)$ and $r>0$, such that for any $x\in \mbb X$, there exists a corresponding $\xi \in \mRd{d}$ with $\mbb K(x, \xi, \theta, r) \subset \mbb X$.} 
	$\mbb U \subset \rCb^{s+1}(\mbb X, \mbb A)$ is homeomorphic to some $\mbb V\subset \mRd{d_v}$ ($d_v<\infty$), such that $\|u(\cdot|v) - u(\cdot|v')\|_{\rCb^{s+1}}\leq |v - v'|$, $\forall v,v'\in \mbb V$. 
\end{assumption}

Denote the \emph{fill distance}:
$$\eta = \eta_{\mbb X\times \mbb U} = \sup_{x\in \mbb X, u\in \mbb U} \min_{i=1,\dots,m} \bra{|x-x_i| + \|u-u_i\|_{\rCb^{s+1}} }.$$
By the above assumption and the smoothness of $f$, the fill distance of $\{f(x_i,u_i)\}_{i=1}^m$ on $\mbb X$ is linearly bounded by $\eta$, and hence we simply identify them. If the sample is deterministically arranged, the fill distance scales down by 
$$ \eta \propto m^{-1/(d_x+d_v)}. $$
Therefore, for any given state-dependent function $g\in \mc{N}_{\kappa}(\mbb X)$, the predicted value at the succeeding instant (namely $\widehat{g(x^+)}=\langle g, \, \hat{T}\hat\phi_{(x,u)} \rangle$), compared to the actual value (namely $g(x^+)= \langle g, \, T\hat\phi_{(x,u)} \rangle$), results in the following error bound. 

\begin{theorem}[Bound on generalized prediction error]\label{th:EDMD.generalization}
	Suppose that Assumptions \ref{assum:1}, \ref{assum:2}, and \ref{assum:3} hold. Then
    \begin{equation}\label{eq:single-step.prediction.error}
        \lvert \widehat{g(x^+)} - g(x^+) \rvert \lesssim \eta^{s-d_x/2} \|T\||x|. 
    \end{equation}
\end{theorem}
\begin{proof}
	It was proved in \cite[Th.~11.17]{wendland2004scattered} that for any region $\mbb X\subset\mRd{d}$ satisfying the interior cone condition, with kernel $\kappa$ for $H^s(\mbb{X})$ ($s=d_x/2+\nu$, $\nu>0$), any $g\in \mc{N}_{\kappa}(\mbb X)$ and its interpolant $s_g$ on points $x_1, \dots, x_m\in \mbb X$ satisfy $|g(x) - s_g(x)| \lesssim \eta_{\mbb X}^\nu \|g\|_{\mc{N}_\kappa} $. 
	\par We use the conclusion in the context of $\rg H^s(\mbb{X})$ instead of $H^s(\mbb{X})$. That is, for any $g\in \rg H^s(\mbb X) \simeq \mc{N}_\kappa(\mbb X)$, $\|g-s_g\|_{\rCb} \lesssim \eta^\nu\|g\|_{\mc{N}_\kappa}$. This inequality is justified by an easily deducible fact that if $g=\sum_{k=1}^{d_x} e_kg_k$ ($g_k\in H^s(\mbb{X})$, then the Sobolev kernel interpolants $s_{g_k}$ constitute the linear--Sobolev interpolant $s_g$, i.e., $s_g = \sum_{k=1}^{d_x} e_ks_{g_k}$. 
    Thus, 
    $$ \lvert \widehat{g(x^+)} - g(x^+) \rvert = \lvert s_g(x^+)-g(x^+) \rvert \lesssim \eta^\nu \|g\|_{\mc{N}_\kappa} |x^+|. $$
    in which we further have 
    $$|x^+|\propto \|\phi_{x^+}\| \leq \|T\|\|\bar\phi_{(x,u)}\| \lesssim \|T\|\|\phi_x\| \propto \|T\||x|.$$
    The proof is completed. 
\end{proof}

In particular, we can let $g=e_1, \cdots, e_{d_x}$ (state component projections) to obtain $\lvert \widehat{x^+}-x^+ \rvert \lesssim \eta^{s-d_x/2}\|T\||x|$. 
That is, the \emph{single-step state prediction error is uniformly proportionally bounded, scaled by $|x|$}. In other words, a \emph{sectorial error bound} in state prediction is guaranteed. 
Such a result is clearly brought by our selection of the linear--radial kernel, with which $\|\phi_x\|$ is proportional to $|x|$. 
Considering the scaling with sample size $m$, we have 
$$\lvert \widehat{x^+}-x^+ \rvert \lesssim m^{-\frac{s-d_x/2}{d_x+d_v}}\|T\||x| .$$ 
The capability of approximating the Koopman--Nemytskii operator therefore strongly depends on the smoothness of the kernel and also the dimensions of the states and the policy space. Smoother systems provide better scaling laws; if $s>d_x+d_v/2$, then the error in proportion to $|x|$ decays at a rate of at least $m^{-1/2}$. 

\begin{remark}[$L^\infty$ versus $L^2$ error]\label{rem:mixed.bound}
    If one is concerned only with the \emph{single-step prediction error}, a mean-squared error ($L^2$-error) can be used as a performance metric. 
    By \cite{kostic2022learning}, the \emph{squared $L^2$-error} can be bounded by $O(m^{-1}\log(1/\delta) + \sqrt{m^{-1}\log(1/\delta)})$ for a confidence of $1-\delta$. 
    This appears to have a better scaling with sample size $m$ if the smoothness index $s$ is low, insensitive to dimensionality due to the irrelevance of fill distance. 
    However, such a non-uniform error bound cannot be easily extended to multi-step predictions. 
    Potentially, a ``reconciled'' error, e.g., uniform \emph{on trajectories} and \emph{mean-square among trajectories}, may be useful for multi-step prediction and policy evaluation. However, a suitable learning formulation to enforce such a bound is lacking. 
\end{remark}
\begin{remark}[Dimensionality reduction and decomposition]
    When the state space and/or policy space have high dimensions, two possible remedies may be helpful to the user. (i) Dimensionality reduction methods can be used to remove the empirically redundant variables. (ii) The system can be possibly decomposed into interconnected subsystems, each having a smaller dimension, so that the learning is performed separately on these subsystems. 
\end{remark} 
\begin{remark}[Online learning and reinforcement learning]\label{rem:online.learning}
    The operator learning here is based on a fixed dataset and hence an offline routine. Due to the pursuit of an $L^\infty$-type error, the error bound relies on the fill distance that does not ``nicely'' scale down, if $s$ is too small or $d_v$ is high.  
    Hence, the user may use an online numerical interpolation or online learning approach to improve the quality of the estimator operator as new data arrives. 
    Potentially, one can iteratively update the Koopman--Nemytskii operator and synthesize the controller (as we will discuss later), conceptually in a reinforcement learning manner to improve the control performance. 
    This is beyond what can be addressed in the current paper. 
\end{remark}

\subsection{Error in multi-step prediction and accumulated cost}\label{subsec:multistep}
Following the previous subsection, if the estimated operator offers a uniform error (scaled by $|x|$) in predicting the succeeding states over a single time step, then the prediction over multiple time steps is anticipated to be correspondingly bounded. 
For any fixed $x\in \mbb X$ and $u\in\mbb U$, let us denote by $\phi_x^0 = \hat{\phi}_x^0 = \phi_x$, and for $t=0, 1, \dots$, denote $\phi_x^{t+1} = T(\phi_x^t \otimes \varphi_u)$ and $\hat{\phi}_x^{t+1} = \hat{T}(\hat{\phi}_x^t\otimes \varphi_u)$. 
\begin{theorem}[Error bound for multi-step prediction]\label{th:multistep.generalization}
	Under the conditions of Theorem \ref{th:EDMD.generalization}, for all $g\in \mc{N}_\kappa(\mbb X)$ and $t=1,2,\cdots$, 
    \begin{equation}\label{eq:multi-step.prediction}
        \left| \ip{g}{\hat{\phi}_x^t - \phi_x^t} \right| \lesssim \eta^{s-d_x/2} t\beta^{t-1} \|g\|_{\mc{N}_{\kappa}} |x|, 
    \end{equation} 
	where $\beta:= \max\{\sup_{u\in \mbb U}\|K_{f_u}\|, \|\hat T\|\}$. 
    If further $\beta < 1$, then 
	$$\left| \ip{g}{\hat{\phi}_x^t - \phi_x^t} \right| \lesssim \eta^{s-d_x/2}\|g\|_{\mc{N}_{\kappa}} |x|, \enspace t\in \mbb{N}.$$
\end{theorem}
\begin{proof}
	We write for any $t=1,2,\cdots$:
	\begin{align*}
		& \left| \ip{g}{\hat\phi_x^{t+1}} - \ip{g}{\phi_x^{t+1}} \right| \\
        & = \left| \ip{Sg}{T(\hat\phi_x^t\otimes \varphi_{u})} - \ip{g}{T(\phi_x^t\otimes \varphi_{u})} \right| \\
        & \leq \left| \ip{(S-\mr{id})g}{T(\hat\phi_x^t\otimes \varphi_{u})} \right| + \left| \ip{g}{T((\hat\phi_x^t - \phi_x^t) \otimes \varphi_{u})} \right| \\
        & \lesssim \eta^{s-d_x/2}\beta^t \|g\| \|\phi_x\| + \left| \ip{K_{f_u}g}{\hat\phi_x^t} - \ip{K_{f_u}g}{\phi_x^t} \right|.
	\end{align*}
    \par The last inequality holds due to the single-step error bound in \eqref{eq:single-step.prediction.error} and the fact that $\|\hat\phi_x^t\|\leq \|\hat T\| \| \hat\phi_x^{t-1} \otimes \varphi_u\| \leq \|\hat{T}\|\|\hat\phi_{x-1}\|$. Hence, starting from any $t\in \mbb N$ and track backwards, we have 
    $$\textstyle \left| \ip{g}{\hat\phi_x^t} - \ip{g}{\phi_x^t} \right| \lesssim \sum_{\tau=0}^{t-1} \eta^{s-d_x/2}\beta^\tau \|K_{f_u}^{t-1-\tau}g\| \|\phi_x\| .$$
    where $\beta^\tau \|K_{f_u}^{t-1-\tau}g\|\leq \beta^{t-1}\|g\|$ and $\|\phi_x\|\propto |x|$. 
\end{proof}

\begin{remark}[Contraction]
	To have a uniform multi-step prediction error, the above theorem provides the condition that the Koopman spectrum lies inside the unit circle in $\mbb C$. 
    As discussed in \cite{tang-ye2025koopman}, such contraction is impossible if the state-RKHS is the Sobolev space $H^s(\mbb{X})$, because the transition on the equilibrium point $\phi_0\mapsto \phi_0$ with $\|\phi_0\|\neq 0$ would result in $\|T\|\geq 1$. 
    Instead, by using the linear--Sobolev space $\rg H^s(\mbb{X})$ associated with the linear--radial kernel (as in this paper) enables contraction, if there is a homeomorphism on the state space $\mbb{X}$ that maps the nonlinear dynamics to the linear dynamics governed by the Jacobian at the origin, and the Jacobian has all eigenvalues inside the unit circle. 
	\par An alternative to the linear--Sobolev space formulation is to adopt the weighted $\mCb$-space (and an associated weighted RKHS) \cite{tang2025koopman}. Specifically, if an exponentially decaying smooth positive function $w(\cdot)$ is known:
	$\inf_{x\in \mbb X} w(f_u(x))/w(x) \leq \alpha < 1$, 
	then we define the \emph{weighted $\mCb$ space}: $C_{\mr{b},w}(\mbb X) = \{w\cdot h: h\in \mCb(\mbb X)\}$, with norm defined by $\|w\cdot h\|_{C_{\mr{b},w}} = \|h\|_{\mCb}$.  
	The Koopman--Nemytskii operator $T$ is then contractive: $\|T\|_{C_{\mr{b},w} \times \mc{N}_\varkappa \rightarrow C_{\mr{b},w}} \leq \alpha < 1$. 
\end{remark}

\par In addition to the bound on multi-step prediction, we also provide a bound on the accumulated cost for \emph{policy evaluation}. In an optimal control language, the stage cost (i.e., the cost at each time in the horizon) is defined by positive functions on the states and input actions, respectively: $q(x) + r(u(x))$, and the terminal cost by a positive function $q_{\mr{f}}(x)$. 
Considering these costs as ``kernel quadratic forms'':
$q(x) = \ip{\phi_x}{Q\phi_x}$, $r(u(x))=\ip{\bar\phi_{(x,u)}}{R\bar\phi_{(x,u)}}$, $q_{\mr{f}} = \ip{\phi_x}{Q_{\mr{f}}\phi_x}$, with $Q$, $Q_{\mr{f}}$, and $R$ being trace-class positive operators, the accumulated cost is expressed as: 
\begin{equation}\label{eq:cost-to-go}
	\textstyle \psi(x, u) = \sum_{t=0}^{\tau-1} \gamma^t \bra{\ip{\phi_x^t}{Q\phi_x^t} + \ip{\bar{\phi}_{x,u}^t}{R\bar{\phi}_{x,u}^t}} + \ip{\phi_x^\tau}{Q_{\mr{f}} \phi_x^\tau} 
\end{equation}
where $\bar{\phi}_{x,u}^t = \phi_x^t \otimes \varphi_u$, $\gamma\in (0, 1]$ is a discount factor. 
We can now consider the difference between $\psi(x, u)$ and its approximation under the data-based estimation:
$$\textstyle \hat{\psi}(x, u) = \sum_{t=0}^{\tau-1} \gamma^t \bra{\langle \hat{\phi}_x^t, Q\hat{\phi}_x^t \rangle + \langle \hat{\bar{\phi}}_{x,u}^t, R\hat{\bar{\phi}}_{x,u}^t \rangle } + \langle \hat\phi_x^\tau, Q_{\mr{f}} \hat\phi_x^\tau \rangle. $$

\begin{theorem}[Error bound for policy evaluation]\label{th:cost-to-go.generalization}
	Under the conditions in Theorem \ref{th:multistep.generalization}, we have 
	\begin{equation}\label{eq:cost-to-go.generalization}
		\begin{aligned}
			&|\hat{\psi}(x,u) - \psi(x,u)|\lesssim \eta^{s-d_x/2}|x|^2  \sum_{t=1}^\tau \gamma^t t(\beta^{t-1}+\beta^{2t-1})
		\end{aligned}
	\end{equation}
	for all $x\in \mbb X$ and $u\in \mbb U$. 
	If further $\beta<1$, then 
	$$\begin{aligned}
	    |\hat{\psi}(x,u) - \psi(x,u)| \lesssim \eta^{s-d_x/2}|x|^2.
	\end{aligned}$$ 
\end{theorem}
\begin{proof}
	By writing the positive operator $Q$ as $Q = \sum_{j=1}^\infty \bar q_j\times \bar q_j$, where $\bar q_j \in \rg H^s(\mbb{X})$, we have 
    $$\textstyle \left| \langle \hat\phi_x^t, Q\hat\phi_x^t \rangle - \langle \phi_x^t, Q\hat\phi_x^t \rangle \right| \leq \sum_{j=1}^\infty \left| \langle \bar q_j, \hat\phi_x^t \rangle^2 - \ip{\bar q_j}{\phi_x^t}^2 \right|. $$
    Due to Theorem \ref{th:multistep.generalization},  
    $| \langle \bar q_j, \hat\phi_x^t \rangle - \langle \bar q_j, \phi_x^t \rangle | \lesssim \eta^\nu \|\bar q_j\| |x|t\beta^{t-1}$, while $| \langle \bar q_j, \hat\phi_x^t \rangle + \ip{\bar q_j}{\phi_x^t} | \leq \|\bar q_j\| (\|\phi_x^t\| + \|\hat\phi_x^t\|) \lesssim \|\bar q_j\|(1+\beta^t)|x|$. 
    Since $\sum_{j=1}^\infty \|\bar q_j\|_{\mc{N}_\kappa}^2 = \mr{tr}(Q)<\infty$, 
    $$\textstyle \left| \langle \hat\phi_x^t, Q\hat\phi_x^t \rangle - \langle \phi_x^t, Q\hat\phi_x^t \rangle \right| \lesssim \eta^\nu |x|^2t\beta^{t-1}(1+\beta^t). $$ 
	In an analogous way we can bound the error between $\langle \hat{\bar{\phi}}_{x,u}^t, R\hat{\bar{\phi}}_{x,u}^t \rangle$, $\langle \bar{\phi}_{x,u}^t, R\bar{\phi}_{x,u}^t \rangle$, as well as the error between $\langle \hat\phi_x^\tau, Q_{\mr{f}} \hat\phi_x^\tau \rangle$ and $\ip{\phi_x^\tau}{Q_{\mr{f}} \phi_x^\tau}$ in the exactly same form. 
\end{proof}

\subsection{Example: liquid storage tank}
To demonstrate the learned Koopman--Nemytskii operator's capability to provide error bounds in single-step state prediction, multi-step state prediction, and policy evaluation as established in the previous subsections, we consider a simple system with $d_x=d_u=1$ --- a liquid tank with a constant-rate inlet stream and a manipulated outlet stream, where the valve position $a\in [-1,1]$ changes the resistance coefficient of the fluid flow in the pipe and thus changes the flow rate that can be delivered by a fixed pump. 
The liquid level of the storage tank, $x$, satisfies the following equation: 
\begin{equation}\label{eq:tank.model}
	x_{t+1}=x_t + 0.2 - \left[11 + 7(1+0.05^a)\right]^{-1/2}.
\end{equation}
Details of the derivation of the model follow from the first principles of fluid mechanics.\footnote{
	Here we consider a tank whose volume is \SI{5}{\cubic\meter} and the inlet stream has a constant flow rate of \SI{0.5}{\cubic\meter\per\minute}. The outlet flow has an adjustable flow rate of $q$ \unit{\cubic\meter\per\minute}. Let the sampling time be \SI{2}{\minute}. Denote the liquid tank storage level as $x$ (ranging from $0$ to $100\%$). The equation is therefore 
	$$x_{t+1} = x_t + (2/5)(0.5-q).$$
	The outlet flow rate is, however, not directly amenable to a controller, but adjusted by a gate valve after a centrifugal pump. The pump has the following characteristic curve: $h = 40 - 44q^2$, where $h$ is the pressure head (in \unit{\meter}), which needs to meet the pressure drop of the outlet pipe (assumed to be \SI{15}{\meter}) in addition to the friction loss. The friction in \unit{\meter} is specified by an ``equivalent length'' $l_e$ in addition to the pipe length $l$ \cite{mccabe1993unit}, i.e., 
	$$\textstyle h = 40 - 44q^2 = 15 + \frac{8\lambda}{\pi^2 g}\frac{l+l_e}{d^5} \frac{q^2}{3600}.$$
	We suppose that the last term above is equal to $28(1+20^{-u})q^2$ with the assumption that the friction of the gate valve is proportional to $20^{-a}$, where $a=-1$, $0$, and $1$ represents when the valve is $1/4$-open, $1/2$-open, and $3/4$-open, respectively. Thus, 
	$$q^2 = 25/[44 + 28(1+20^{-a})].$$
	The model thus becomes \eqref{eq:tank.model} after translating the equilibrium point to $0$.
}

\par Let $\mbb X = [-2, 2]$ on which we assign the Wendland-form Sobolev kernel $\kappa_{\mr{Sob}}=\kappa_{d_x, k}^{\mr{Wen}}$ such that $\mc{N}_{\kappa_{d_x, k}^{\mr{Wen}}}(\mbb X) \simeq H^{(d+1)/2+k}(\mbb X)$. We use $k=1$ and hence $s=2$. 
Hence, $\kappa = \kappa_{\mr{lin}}\kappa_{d_x, k}^{\mr{Wen}}$ renders $\mc{N}_{\kappa}(\mbb X) \simeq \rg H^2(\mbb X)$. (See Footnote \ref{fn:Wendland} on the Wendland kernels.)
We tune $\sigma_x$ so that the kernel matrix has a sparsity below $0.5$. 
Let $\mbb A = [-1, 1]$ and $\mbb U = \{x\mapsto \tanh(kx): k=10^v, \, v \in [-1, 1] \}$, on which the kernel is defined as the radial Wendland kernel $\varkappa = \kappa_{d_u, k}^{\mr{Wen}}$, and $\sigma_u$ is tuned in the same way. 

\begin{figure}[!t]
\centering
	\includegraphics[width=0.75\columnwidth]{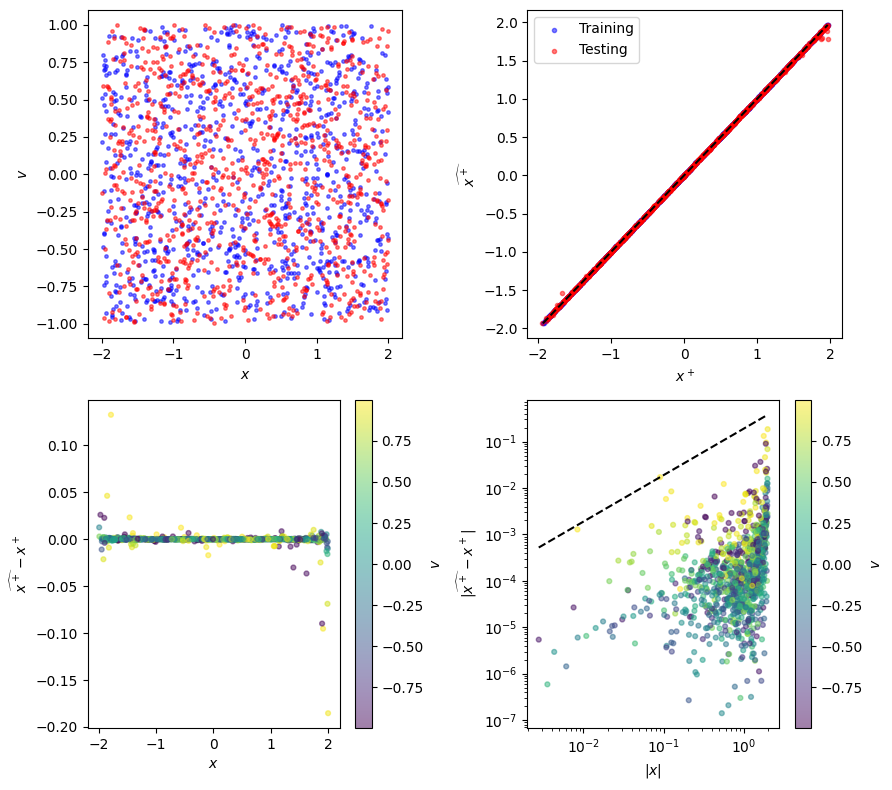}
	\caption{Single-step state prediction error of the estimated Koopman--Nemytskii operator. Subplot (1, 1): training and testing data points. Subplot (1, 2): predicted succeeding state $\widehat{x^+}$ versus true $x^+$. Subplot (2, 1): prediction error plotted against $x$. Subplot (2, 2): prediction error in logarithmic scale.}\label{fig:tank-singlestep}
    \vspace{-1.0em}
\end{figure}
\par Via kernel EDMD as described in \S\ref{subsec:kernel.DMD}, the estimated Koopman--Nemytskii operator $\hat{T}$ is obtained via $m=1000$ training data points, which well cover the region $\mbb{X}\times \mbb{V}$. Using it under any new $(x,u)$-pair, we are able to predict the succeeding state. Its comparison with the true succeeding state shows low error (with a RMSE of $9.16\times 10^{-3}$). 
The plot of state prediction error $|\widehat{x^+} - x^+|$ versus $|x|$ confirms that the error is proportional to state (see Fig. \ref{fig:tank-singlestep}). Based on the testing sample, we estimate it as $|\widehat{x^+}-x^+|\leq 0.189|x|$. This observation is consistent with the claim of Theorem \ref{th:EDMD.generalization}. 

\begin{figure}[!t]
\centering
	\includegraphics[width=0.75\columnwidth]{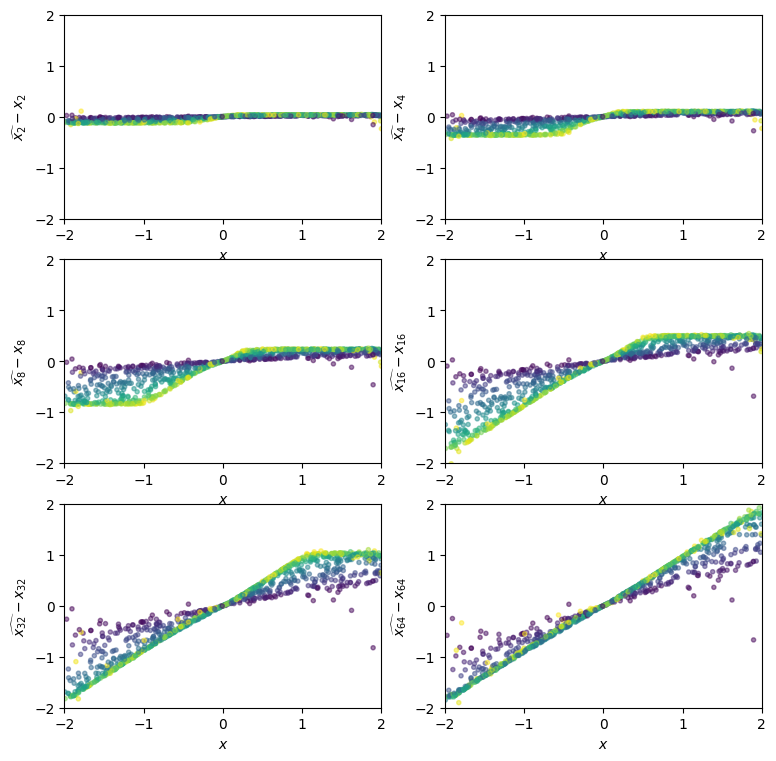}
	\caption{Multi-step state prediction error of the estimated Koopman--Nemytskii operator for $t=2, 4, 8, 16, 32, 64$.}\label{fig:tank-multistep}
    \vspace{-1.0em}
\end{figure}
\par In the afore-mentioned controller parameter range, the system is always closed-loop stable. We hence apply the estimated operator $\hat{T}$ under each test data point $(x,u)$ to obtain the $\hat{\phi}_x^t$ under the policy $u$ starting from $x$, for a range of $t$.
For $t=2, 4, 8, 16, 32, 64$, the state prediction errors are plotted against the \emph{initial state} $|x|$ with color map on $u$, as shown in Fig. \ref{fig:tank-multistep}. 
As the time progresses, the error propagation results in a lowering prediction precision, although, as Theorem \ref{th:multistep.generalization} correctly claims, that the error $|\widehat{x}_t - x_t|$ must remain proportionally bounded by the initial state $|x|$, and that proportion remains bounded. 

\par
\begin{figure}[!t]
	\centering
	\includegraphics[width=0.75\columnwidth]{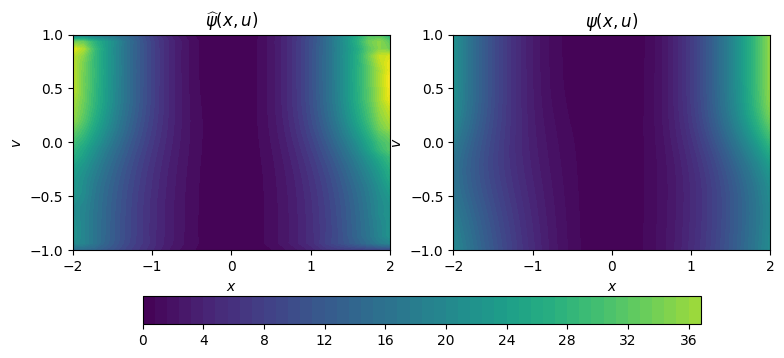}
	\caption{Accumulated cost and its prediction under the Koopman--Nemytskii operator.}
	\label{fig:tank-evaluation}
    \vspace{-1.0em}
\end{figure}

 We then consider the approximation of the cost accumulated in a horizon of $24$ time instants (namely $48$ minutes):
$$\textstyle \psi(x, u) = \sum_{t=0}^{24} \gamma^t\bra{x_t^2 + a_t^2} = \sum_{t=0}^{24} \gamma^t \bra{x_t^2 + u(x_t)^2}, $$
where $\gamma=0.85$.
The comparison of the actual cost $\psi$ and the predicted cost $\hat{\psi}$ under $\hat{T}$ are plotted against $(x, v)$-combinations in Fig. \ref{fig:tank-evaluation} under the kernel EDMD approximation. 
As can be observed, the predicted costs are close to the actual values; the errors enlarge as $x$ is closer to $\pm 2$ and as $v$ becomes larger (i.e., under higher controller gain parameters), but vanish as $x\to 0$ as predicted by Theorem \ref{th:cost-to-go.generalization}. 

\section{Feedback Controller Synthesis}\label{sec:Experiments}
Now let us return to the rationale for proposing an operator that takes in the canonical features of state and \emph{policy} for prediction and policy evaluation. By using the policy feature instead of a possible input action feature as conceptualized in \eqref{eq:conceptual.input-state.viewpoint}, in its definition, the Koopman--Nemytskii operator avoids the composition of the input action with the policy. 
Hence, in controller synthesis, i.e., when monolithically finding an optimal feedback law within a parametric policy class, the search can be considered as a \emph{search over the policy features}, without using the composition operation to relate the policy to its performance evaluation. 

\subsection{Objective formulation}
Starting at any given initial state $x\in \mbb X$, the performance of a policy $u\in \mbb U$ is evaluated by the accumulated cost over an infinite horizon:
\begin{displaymath}
    \textstyle \psi_\infty(x, u) = \sum_{t=0}^\infty \gamma^t\left[ q(x_t) + r(u(x_t)) \right], 
\end{displaymath}
where $x_0 = x$, $x_{t+1} = f(x_t, u(x_t))$ ($t\in \mN$). As in \S\ref{subsec:multistep}, we assume that the stage cost can be expressed as a ``kernel quadratic form'', i.e., $q(x_t) = \ip{\phi_x^t}{Q\phi_x^t}$ and $r(u(x_t)) = \ip{\bar\phi_{x, u}^t}{R\bar\phi_{x,u}^t}$ with $\bar\phi_{x,u}^t = \phi_x^t \otimes \varphi_u$. 
The evaluation of the policy is thus taken as the average over a probability distribution of states:
$$ J_\infty(u) = \mbb{E}_{x\sim \mbb{P}} \Bra{\psi_\infty(x,u)}. $$

\par While $J_\infty(u)$ may not be exactly evaluated at any $u$, due to the possibly continuous distribution $\mbb P$ and the infinite horizon in $\psi_\infty$, there always exists a discrete distribution $\hat{\mbb{P}}$ supported on finitely many points $\{x_1, \cdots, x_{m_x'}\}$ and a sufficiently long horizon $\tau\in \mN$ such that 
\begin{displaymath}
    \begin{aligned}
        \mbb{E}_{x\sim \hat{\mbb{P}}} \Bra{\psi_\tau(x,u)} := \frac{1}{m_x'}\sum_{i=1}^{m_x'} \Bra{\sum_{t=0}^{\tau-1} \gamma^t\left[ q(x_{i,t}) + r(u(x_{i,t})) \right] + q_{\mr{f}}(x_{i,\tau})} 
    \end{aligned}
\end{displaymath} 
with a terminal cost $q_\mathrm{f}$, is a precise estimation of $J_\infty(u)$ with an arbitrarily small uniformly bounded error. This is due to the continuity of $q$ and $r$ in addition to the boundedness of state space $\mbb X$ and action space $\mbb A$. 
In a data-driven setting, the evaluation of $\psi_\tau(x_i, u)$ can be approximately performed by the Koopman--Nemytskii operator estimated from kernel EDMD. As proven in Theorem \ref{th:cost-to-go.generalization}, the estimated $\hat \psi_\tau(x_i)$ has an error proportional to $|x|^2$ (and hence uniform over $\mbb{X}$), which is further independent of $\tau$ if contraction holds, i.e., if $\beta := \max\{\|\hat T\|, \sup_{u\in \mbb U}\|K_{f_u}\|\}<1$. 
Thus, we denote
\begin{equation}\label{eq:synthesis.objective}
    J(u) = \mbb{E}_{x\sim \hat{\mbb{P}}} \Bra{\hat\psi_\tau(x,u)} 
\end{equation} 
as the practically computable, Koopman--Nemytskii learning-based policy evaluation. Therefore, the problem of minimizing $J(u)$ over $u\in \mbb U$ is actually of interest. 
For any precision $\epsilon>0$, we have sufficiently large $m_x'$, $\tau$, and $m$ (in kernel EDMD), such that as long as the optimizer $u^\star$ of $\hat J(u)$ is found, we have $J_\infty(u^\star) - J(u^\star) < \epsilon$. 

\begin{proposition}[Regularity of the objective function]
    Suppose that $q(x_t) = \ip{\phi_x^t}{Q\phi_x^t}$ and $r(u(x_t)) = \ip{\bar\phi_{x, u}^t}{R\bar\phi_{x,u}^t}$ with $Q$ and $R$ being trace-class operators in $\mc{N}_{\kappa}(\mbb X)$ and $\mc{N}_{\bar\kappa}(\mbb X\times \mbb U)$, respectively. Let $\varkappa$ be such that $\mc{N}_\varkappa(\mbb U)\simeq H^s(\mbb U)$. Then $J\in \mc{N}_\varkappa(\mbb U)$. 
\end{proposition} 
\begin{proof}
    By decomposing $Q = \sum_{j=1}^\infty \bar q_j \times \bar q_j$, we have $q(x_t) = \sum_{j=1}^\infty \bar q_j(x_t)^2$, where each $\bar q_j\in \rg H^s(\mbb X)$. The same decomposition can be written for $R$. Hence, it only remains to prove that each $x_t$ ($t\in \mN$) is an $H^s$-class of $u$, since it would justify that $\bar q_j(x_t)^2$ and $\bar r_j(x_t, u_t)^2$ are in the $H^s$ class as a function of $u$. This is a direct implication of the $C^s$ property of $f$. 
\end{proof}
Hence, the problem is to minimize an objective function that resides in an RKHS, while it may be nonconvex and lacks an explicit algebraic expression, although it can be evaluated on sample points.

\subsection{Kernel-based optimization algorithm}
Conceptually, one can reformulate the global optimization of a nonconvex function as the optimization of probability distributions on $\mbb U$, i.e., $\min_{\mbb Q} \int J(u)\mr{d}\mbb{Q}(u)$, so that the optimal solution corresponds to any distribution supported on global optima. The reformulated problem is indeed convex, but should be interpreted as an \emph{infinite-dimensional linear programming problem}. If the objective is polynomial, the problem is to optimize the moments of the distribution, and the dual problem is to find the maximum constant $c$ such that $J-c$ is a nonnegative polynomial. The resulting momentum/polynomial optimization problem allows a disciplined solution via hierarchical finite-dimensional approximations \cite{henrion2020moment, nie2023moment}. 

\par Generalizing to a non-polynomial RKHS setting, the formulation of Rudi et al. \cite{rudi2025finding} considers $J-c$ as a nonnegative kernel quadratic form, i.e., a kernel sum-of-squares form, and enforces this constraint on some sample points $\{u_1, \cdots, u_{m_u'}\}$:
\begin{equation}\label{eq:rudi} 
\begin{aligned}    
    \max_{c\in \mR, \, A\in \mc{S}_+} \enspace & c-\lambda \operatorname{tr}A \\
    \text{s.t.} \quad & J(u_i)-c = \ip{\varphi_{u_i}}{A\varphi_{u_i}}, \, i=1,\cdots,m_u'.
\end{aligned}
\end{equation}
where $\mc{S}_+$ stands for the space of positive self-adjoint trace-class operators on $\mc{N}_\kappa(\mbb U)$. Since the constraints are on sample points only, it suffices to find a positive semidefinite matrix $B$ such that 
$$J(u_i)-c = g_i^\top Bg_i,\, i=1,\cdots,m_u'$$
where $g_i$ is the $i$-th column of the Gram matrix $G_\varkappa = [\varkappa(u_i, u_j)]$, and replace $-\lambda\operatorname{tr}A$ with $-\lambda\operatorname{tr}B$. The regularization parameter $\lambda>0$ serves the purpose of leveraging the smoothness of the kernel --- when $\lambda=0$, the optimized $c$ is simply the sample minimum regardless of the ``complexity'' of $J-c$ as a quadratic form, while using $\lambda>0$ penalizes such a complexity and thus constrains the variation of $J-c$ from the sample points. 
The accuracy of \eqref{eq:rudi} was proved in \cite[Th.~6]{rudi2025finding}, which we translate into our setting as below. 
\begin{theorem}[Accuracy of kernel-based optimization]\label{th:kernel.optimization}
    Suppose that (i) the controller parameter set $\mbb V = \cup_{v\in \mbb{V}_0} \mbb{B}(v, r)$ for some $\mbb V_0\subset \mbb V$ and $r>0$, (ii) $s > d_v/2 + 2$ and $\mc{N}_\varkappa(\mbb U) \simeq H^s(\mbb V)$, (iii) $\mbb V$ contains finitely many global optima, all of which are isolated points and not on the boundary of $\mbb V$, and at which the Hessian matrices are all positive definite. 
    Then, with sufficiently large $m_u'$ and $\lambda \gtrsim (\frac{1}{m_u'} \log \frac{2^{d_u}m_u'}{\delta})^{2/d_v}$, with probability $1-\delta$ over the random choice of $\{u_1, \cdots,u_{m_u'}\}\subset \mbb U$, it is guaranteed that $ |c - \min_{u\in \mbb{U}} J(u)| \lesssim \lambda$. 
\end{theorem}

To enable finding the location of the optimum $u^\star$, if the global optimum is unique, then a refined problem can be solved, wherefrom a parabola centered at $(\bar v, 0)$ with curvature $\nu$ is determined to be a lower estimate of $J-c$:
\begin{equation}\label{eq:rudi.2}
\begin{aligned}
    & \max_{c\in \mR, \, \bar v\in \mR^{d_v}, \, B\succeq 0} \enspace c- \frac{\nu}{2}|\bar v|^2 -\lambda \operatorname{tr}B \\
    & \text{s.t.} \enspace J(u_i)-c-\frac{\nu}{2}|v_i|^2 + \nu v_i^\top \bar{v} = g_i^\top Bg_i, \, i=1,\cdots,m_u'.
\end{aligned}
\end{equation}
The problem is feasible, provided that $\nu$ is small enough to underestimate the Hessian curvature at the optimum. Thus, the optimized $\bar v$ is approximately the optimal controller parameter $v^\star$ corresponding to $u^\star$.\footnote{As a high-dimensional semidefinite program, \eqref{eq:rudi.2} is computationally expensive to solve in its primal form. This work follows the suggestion in \cite{rudi2025finding} to solve the dual problem by a damped Newton algorithm. See \cite[Sec.~6]{rudi2025finding} for details.}

\subsection{Example: A chemical reactor}
We consider the following stirred tank reactor model with $2$ states and $1$ input action. 
Its continuous-time model\footnote{Physically, $a$ is the throughput flow rate, $x_1$ is an intermediate chemical concentration, and $x_2$ is the product concentration, all in relative deviation ratios from their steady-state values. The model comes from \cite[Prob.~7.7]{kravaris2021understanding}.} is written as $\dot{x} = f_0(x)+f_1(x)a$ in which
$$ f_0(x) = \begin{bmatrix}
    \frac{3-x_1}{4} - \frac{9(1+x_1)}{4(3+2x_1)} \\
    -\frac{3(1+x_2)}{4} + \frac{9(1+x_1)}{4(3+2x_1)}
\end{bmatrix}, \enspace 
f_1(x) = \begin{bmatrix}
    \frac{3-x_1}{4} \\
    -\frac{1+x_2}{4}
\end{bmatrix},$$
and we use a sampling time of $0.2$. The state space $\mbb X = [-1/4,1/4]$. 
The policy space $\mbb U$ comprises of two-parameter linear feedback laws $\mbb U = \{x\mapsto v_1x_1+v_2x_2: v_1, v_2\in [-1, 1]\}$. This range of controller parameters is based on the optimal LQR controller gain $[-0.546, 0.050]$, with respect to a cost of $\int_0^\infty (x(t)^2+a(t)^2)\dif{t}$. 
The Sobolev kernels are in the Wendland form. In concert with the optimization algorithm in the previous subsection that requires $s>d_x/2+2$, we use the Wendland kernel with order parameter $k=2$. 

\par
\begin{figure}[!t]
	\centering 
    \includegraphics[width=0.75\columnwidth]{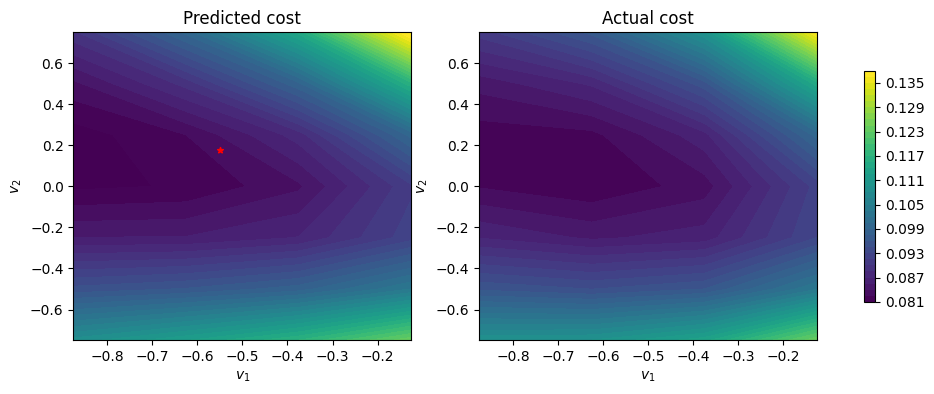}
    \includegraphics[width=0.75\columnwidth]{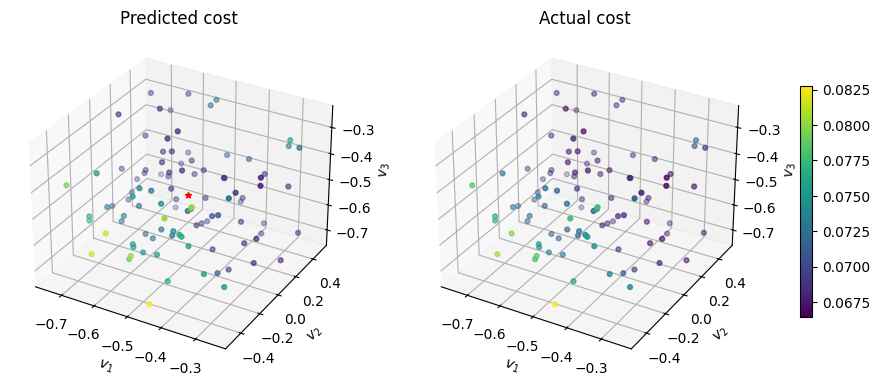}
	\caption{Cost function evaluated for various policies and the optimal policy determined by kernel-based optimization, for 2-parameter policies (upper) and 3-parameter policies (lower).}
	\label{fig:kravaris-cost}
    \vspace{-1.0em}
\end{figure}

 Following the kernel EDMD estimation procedure, we obtain an approximate Koopman--Nemytskii operator with $1000$ training data points. 
Applying the learned operator to predict the states in $25$ sampling times, starting from $m_x'=50$ randomly selected initial states in $\mbb X$, the estimated cost can be calculated for any policy in $\mbb U$. This cost evaluation is performed for a grid of controller parameters $v_1 \in [-0.75, 0]$ and $v_2 \in [-0.75, 0.75]$, using $q(x) = x^2$, $r(a) = a^2$, $\gamma = 1$, and $q_{\mr{f}}(x) = 3x^2$ at $\tau = 25$. 
Fig. \ref{fig:kravaris-cost} shows the contour plots for the predicted costs $\hat J(u)$ in comparison to the actual costs (evaluated by the average under same initial states). It can be seen that the profiles appear similar. By calling the kernel-based optimization algorithm \eqref{eq:rudi.2}, we obtain an empirical optimum $v^\star = (-0.549, 0.175)$, as marked by the star point on the first subplot of Fig. \ref{fig:kravaris-cost}. 

\par In addition, we consider the possibility to enrich the family of control laws to three-parameter policies:  
$\mbb U = \{u:x\mapsto v_1x_1 + v_2x_2 + v_3 \frac{3x_1}{3+2x_1}\}$. The same Koopman--Nemytskii operator learning and prediction routines based on the learned operator can be performed. 
The predicted and actual cost profiles, now dependent on three controller parameters, are shown as the lower two subplots in Fig. \ref{fig:kravaris-cost}. 
Kernel-based optimization suggests $v^\star = (-0.502, 0.0654, -0.506)$. 

\par
\begin{figure}[!t]
    \centering
    \includegraphics[width=0.75\columnwidth]{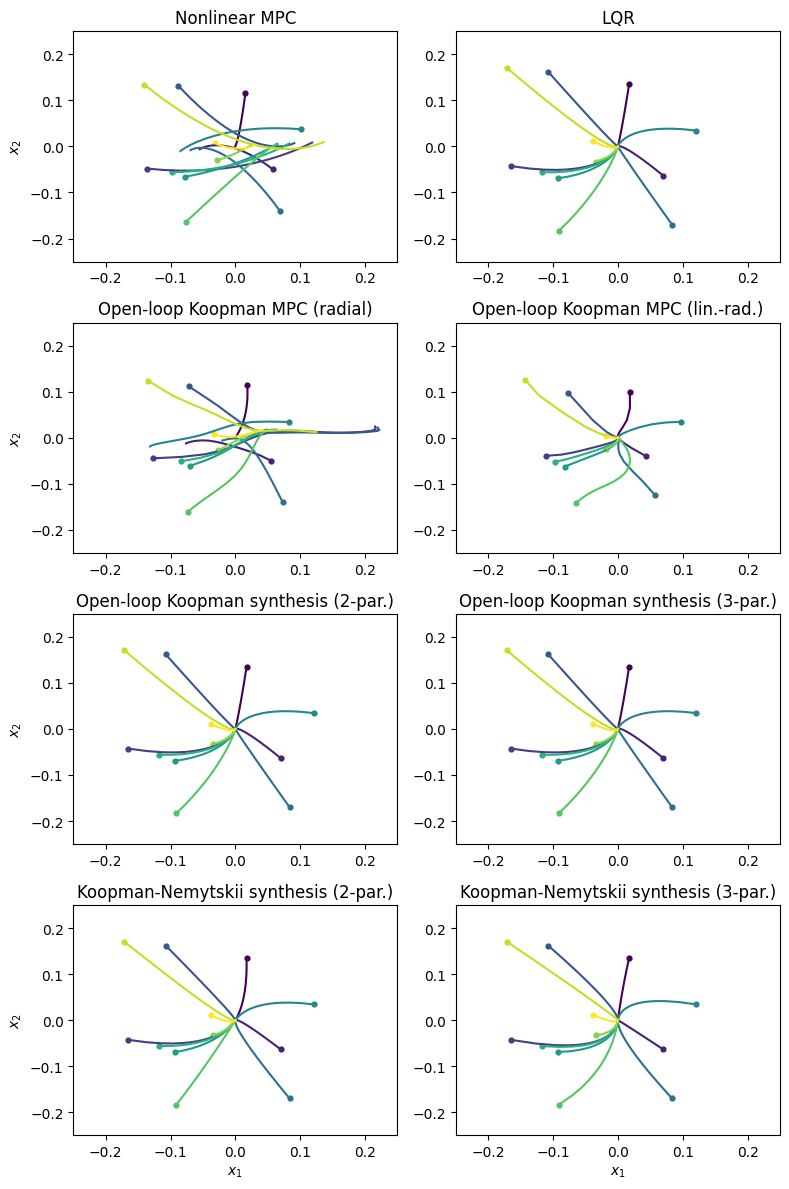}
    \caption{Comparison of state orbits among control strategies.}
    \label{fig:kravaris-controller-comparison}
    \vspace{-1.0em}
\end{figure}

 The synthesized two ``Koopman--Nemytskii controllers'' (2-parameter and 3-parameter) are compared to the following six alternative control strategies. 
\begin{enumerate}
    \item Based on the true model, a nonlinear MPC with the same sampling time and the same $q$, $r$, and $q_{\mr{f}}$ that define the stage and terminal costs is deployed. This strategy assumes that the true dynamics is known \textit{a priori}. 
    \item Assuming that the linearized model at the origin is known, the LQR controller as an explicit feedback law is obtained. 
    \item The ``input--state operator'' \eqref{eq:conceptual.input-state.viewpoint} with the radial Sobolev kernel is learned with kernel EDMD and hence generates a state prediction model, based on which a nonlinear MPC is deployed. This is exactly the strategy of \cite{bold2025kernel}, where it was suggested to add the equilibrium point $(x=0, a=0)$ to the dataset. 
    \item In the previous strategy, the Sobolev kernel is replaced by the linear--radial kernel. These two strategies are referred to hereafter as ``\emph{open-loop Koopman MPC}''. 
    \item We can use the ``input--state operator'' as the open-loop surrogate dynamical model, composed with any given parameterized policy to evaluate the policy's cost. Such evaluations are applied to some sampled policies for kernel-based optimization. 
    The difference from the proposed method is that the operator model does not directly link the policy to the state transitions, but only the input actions. 
    A two-parameter controller is obtained at $v^\star = (-0.500, 0.200)$. We call this approach as ``\emph{open-loop Koopman synthesis}''. 
    \item In the previous strategy, the three-parameter structure can be alternatively used. The optimal controller parameters are found as $v^\star = (-0.485, -0.018, -0.539)$. In these two open-loop Koopman synthesis strategies, we use the linear--radial kernel. 
\end{enumerate}
These controllers are simulated under $12$ randomly picked initial states for a duration of $50$ sampling times, and query their total costs $\sum_{t=0}^{50} (x_t^2 + a_t^2)$ (geometrically averaged over $12$ orbits) as well as the total computational time needed for repeated MPC optimization (algebraically averaged).\footnote{The computations are performed on a MacBook Pro with 14 cores (10 performance and 4 efficiency. Codes are written in Python 3.13.2. MPC is set up with the \texttt{do-mpc} library.} 
These performance metrics are summarized in Tab. \ref{tab:control.comparison}, and the resulting closed-loop orbits are shown in Fig. \ref{fig:kravaris-controller-comparison}.

\begin{table}[!t]
    \centering\small
    \begin{tabular}{c|c|c}
        \hline
        Strategy & Cost & Time \\
        \hline
        Nonlinear MPC & $0.236$ & \qty{296}{\ms} \\
        LQG & $0.065$ & \qty{13.8}{\us} \\
        \hline 
        Open-loop Koopman MPC (radial) & $0.309$ & \qty{155}{\s} \\
        Open-loop Koopman MPC (lin.--rad.) & $0.028$ & \qty{195}{\s} \\
        \hline 
        Open-loop Koopman synthesis (2-par.) & $0.065$ & \qty{13.7}{\us} \\
        Open-loop Koopman synthesis (3-par.) & $0.065$ & \qty{10.5}{\us} \\
        \hline 
        Koopman--Nemytskii synthesis (2-par.) & $0.066$ & \qty{14.4}{\us} \\
        Koopman--Nemytskii synthesis (3-par.) & $0.072$ & \qty{20.6}{\us} \\
        \hline
    \end{tabular}
    \caption{Comparison of control strategies by accumulated cost and computational time.}
    \label{tab:control.comparison}
    \vspace{-1.0em}
\end{table}

\par We have the following observations. 
\begin{itemize}
    \item As expected, the computational time needed by offline synthesized controllers with explicit laws is negligible vis-{\`{a}}-vis MPC strategies. 
    Using a kernel-based operator model in MPC is especially computationally cumbersome, since its state prediction model in the MPC is not only nonlinear, but also contains hidden states that are associated with the training data, which are further repeated throughout the prediction horizon (thus of number $m\tau = 25000$). 
    From this point of view, it is more desirable to pursue a synthesis method than MPC. 
    \item The achievement of asymptotic stability primarily depends on the capability of capturing local dynamics around the equilibrium point, which requires the use of a linearized Jacobian model or a linear--radial kernel-based operator model. 
    Even for true model-based nonlinear MPC (without using terminal constraints or input bounds), the performance is lost. As this should be attributed to the non-globality of nonlinear MPC solver, from a practical point of view, using a linear--radial kernel provides more robust performance. 
    \item Both the learned open-loop Koopman operator (under the linear--radial kernel), composed with any assigned parametric class of policies, and the Koopman--Nemytskii operator directly learned on this policy class, can be used for state prediction and hence policy evaluation. Because of this, the synthesized controllers via strategies 5/6 and strategies 7/8 are similar. 
    However, in the ``open-loop Koopman synthesis'' approach, it is noteworthy that (i) the regularity conditions for the existence of Koopman--Nemytskii operator are already tacitly assumed to hold, (ii) the information used for policy evaluation essentially does not differ from that used for Koopman--Nemytskii policy evaluations, and (iii) the kernel-based optimization still relies on policy parameterization and a policy kernel. 
    \emph{By introducing the notion of Koopman--Nemytskii operator and its learning, the policy evaluation and optimization procedures are explicated.}
\end{itemize}

\section{Conclusion}\label{sec:Conclusion}
\par In this paper, a Koopman-like linear operator representation of nonlinear controlled systems, named Koopman--Nemytskii operator, is proposed, and its data-based estimation, generalization errors, and use for controller synthesis are discussed. 
Under regularity conditions, the \emph{Koopman--Nemytskii operator} is a well-defined, bounded linear operator from the tensor product of an RKHS of states and a RKHS on the space of feedback laws to the foregoing RKHS, mapping the canonical features of a state and the canonical feature of a policy to that of the succeeding state. 
As such, one-step or multi-step state predictions, as well as the prediction of accumulated cost under control, can be performed. In particular, the prediction by the Koopman--Nemytskii operator estimated under kernel EDMD, provided sufficiently small fill distance, is found to give bounded errors scaled with $|x|$. This proportionally bounded error further depends on the dimensions of the state and policy spaces as well as the kernel smoothness that fits the system's regularity. Per discussions in \S\ref{subsec:kernel.DMD}, the estimation error is in the order of $O(m^{-\nu/d})$ (assuming a smoothness index of $d_x/2+\nu$ and $d$ is the total dimension of state and policy parameters). 
Practically, one may resort to auxiliary reduction, decomposition, or online learning approaches to alleviate the curse of dimensionality, if the smoothness index $s$ is low. 

\par To the end of controller synthesis, based on the RKHS concept, the paper adopts a \emph{kernel-based optimization} formulation over the policy space. The problem is solvable in a data-driven (sample-based) manner that involves policy evaluation, using multi-step predictions based on the learned Koopman--Nemytskii operator. 
At this point, it remains an open question whether it is possible to rigorously \emph{synthesize controllers over an unrestricted nonparametric class} based on some operator model, although such a model essentially treats the nonlinearity in a black-box linearity in infinite dimensions. 
For discrete-time systems, where input affinity cannot be expected, our formulation is yet restricted to a prespecified parametric policy class. 
The author hypothesizes that for continuous-time input-affine systems, convex formulations for controller synthesis based on semigroup notions should exist. This is left for a separate work. 

\bibliographystyle{ieeetr}
\bibliography{mybib.bib}

\end{document}